\documentclass[preprint,12pt]{elsarticle}

\usepackage{lineno,hyperref}
\modulolinenumbers[5]

\journal{Nuclear Physics A}

\usepackage{graphicx} % check these packages in TeX Live for NPA
\usepackage{amssymb,amsmath}
\usepackage{bm}

\begin{document}

\begin{frontmatter}

\title{Hidden dibaryons in one- and two-pion production \\ in $NN$ collisions}

%\tnoteref{mytitlenote} --- put inside \title at its end if needed
%\tnotetext[mytitlenote]{Fully documented templates are available
%in the elsarticle package on
%\href{http://www.ctan.org/tex-archive/macros/latex/contrib/elsarticle}{CTAN}.}

% \href doesn't work!

\author[mymainaddress]{M.N. Platonova\corref{mycorrespondingauthor}}
\cortext[mycorrespondingauthor]{Corresponding author.}
\ead{platonova@nucl-th.sinp.msu.ru}

\author[mymainaddress,mysecondaryaddress]{V.I. Kukulin}
%\ead{kukulin@nucl-th.sinp.msu.ru}

\address[mymainaddress]{Skobeltsyn Institute of Nuclear Physics, Lomonosov Moscow State University, Leninskie Gory 1/2, 119991 Moscow, Russia}

\address[mysecondaryaddress]{Institut f\"{u}r Theoretische Physik, Universit\"{a}t T\"{u}bingen, Auf der Morgenstelle 14, D-72076 T\"{u}bingen, Germany}

\begin{abstract}
Processes of one- and two-pion production in $NN$ collisions are
considered in connection with excitation of intermediate dibaryon
resonances. In particular, relative contributions of the
conventional meson-exchange and dibaryon excitation mechanisms in
the reaction $pp \to d \pi^+$ are investigated in detail.
Inclusion of the intermediate isovector dibaryon resonances is
shown to essentially improve the description of experimental data
for this reaction, provided the soft meson-baryon form factors
consistent with $\pi N$ elastic scattering are used. Manifestation
of the intermediate isoscalar and isovector dibaryons in the
two-pion production processes is also studied. The role of the
isovector dibaryon resonances in the reaction $pp \to pp \pi\pi$
is discussed for the first time. An explanation of the observed
strong differences between two-pion production cross sections in
$pn$ and $pp$ collisions based in part on the analysis of dibaryon
structure is suggested.
\end{abstract}


\end{frontmatter}

%\linenumbers

\section{Introduction: Dibaryons are ``to be or not to be''?}
%\paragraph{}
\label{intro} Search for dibaryon resonances and their
manifestations in hadronic and electromagnetic processes is a
long-standing problem which takes its origin in the late 1970ies
(see the basic Refs.~\cite{Jaffe77,Mulders78,Aerts78,Matveev77}
and also reviews~\cite{Makarov82,Locher84,Strak91}). In that time,
the first experimental indications appeared for existence of the
number of dibaryon states. In particular, in elastic scattering of
polarized protons $\vec{p}+\vec{p}$, the signals of a whole series
of isovector dibaryons in $^1D_2$, $^3F_3$, $^1G_4$, etc., partial
waves were found~\cite{Auer77,Auer78,Biegert78} (see also later
experimental works~\cite{Auer82,Bertini88,Auer89}). Besides that,
rather convincing though indirect indications for an isoscalar
dibaryon with quantum numbers $I(J^P) = 0(3^+)$ were found in
measurements of the outgoing proton polarisation in the deuteron
photodisintegration process $\gamma d \to pn$ at energies
$E_{\gamma}\simeq
400$--$600$~MeV~\cite{Kamae77,Kamae77-2,Ikeda80}.

It is worth emphasizing that the first theoretical prediction of
dibaryon states based on SU(6) symmetry was done still in 1964 in
the pioneering work of Dyson and Xuong~\cite{Dyson64}, appeared
only several months after Gell-Mann's first publication on the
quark model of hadrons~\cite{GellMann64}. In the following, a
number of dibaryon states were also predicted and investigated
within the QCD-inspired
models~\cite{Mulders80,Kondr87,Simonov81,Konno87,Garc97}. At the
same time, a series of theoretical works appeared (see,
e.g.,~\cite{Kamo79,Kanai79,Ferreira83,Barannik86}) where dibaryon
degrees of freedom (d.o.f.) in different hadronic processes were
considered. In those works, however, the dibaryon parameters were
adjusted \emph{ad hoc} to describe the observables of a particular
process under consideration, with no explicit relation to both
microscopic quark models and the description for other types of
hadronic processes, where the same dibaryon resonances might
participate.

From the other hand, it was
demonstrated~\cite{Simonov79,Niskanen78,Niskanen79,Mizutani81,Lamot87,Grein84}
that the basic features of some hadronic processes, such as $\pi d
\to \pi d$, $NN \leftrightarrow \pi d$, etc., where the claimed
dibaryon resonances were expected to manifest themselves, can be
described within the framework of conventional meson-exchange
mechanisms without any dibaryon contributions. Hence, it turned
out to be very difficult to draw some definite conclusions about
existence (or absence) of dibaryon resonances and their role in
hadronic processes. The situation was worsened by the fact that,
in spite of extensive searches, in that time (in 1980--90ies) no
quite convincing experimental evidences for dibaryons were
found~\cite{Seth88}. As a result, the general interest to the
problem faded away.

Recently, however, the situation around dibaryons began to change
rapidly. A number of new inspiring results have appeared, that
have led to some kind of renaissance in the area of dibaryon
physics. One of such results has been the prediction of a strange
$H$-dibaryon in the lattice QCD calculations (see,
e.g.,~\cite{Shanahan11,Carames13}) and the following initiation of
a big experimental program at JPARC~\cite{Sako14} aimed at
searching for the $H$-dibaryon. It is worth mentioning that
unsuccessful experimental search for this dibaryon in previous
years (since its first prediction by Jaffe~\cite{Jaffe77})) was
one of the reasons for scepticism against the existence of
dibaryons (see, e.g.,~\cite{Seth88,Bugg14}).

The second not less important result is related to the non-strange
dibaryons. It is the recent experimental finding of the isoscalar
dibaryon resonance $\mathcal{D}_{03}(2380)$ with $I(J^P) = 0(3^+)$
(first predicted by Dyson and Xuong~\cite{Dyson64}) in the
two-pion production reactions $pn \to d \pi\pi$, $dd \to
{}^4\rm{He} \pi\pi$ and $pd \to {}^3\rm{He}
\pi\pi$~\cite{Bash09,Adl11,Adl13-iso,Adl12,Adl15} and an explicit
relation of this resonance to the well-known
Abashian--Booth--Crowe (ABC) effect~\cite{Abashian60,Booth61},
i.e., an anomalous enhancement in the cross sections of these
reactions just above the two-pion threshold. Although the
interrelation between this dibaryon formation in the
$2\pi$-production reactions and the ABC effect was predicted
already in an old paper~\cite{Kamae77-2}, the reliable
experimental data that confirmed this prediction have appeared
only 30 years later. This has become possible mainly due to
considerable progress achieved in experimental technique
(simultaneous registration of three and more particles in
coincidence, measurements in full $4\pi$ geometry, etc.).

The remarkable features of the $\mathcal{D}_{03}(2380)$ resonance
are its mass, lying much (80 MeV) lower than the threshold of
simultaneous excitation of two $\Delta$ isobars, and its rather
narrow width $\Gamma_{\mathcal{D}_{03}} \simeq 70$~MeV. Just these
features made it possible to separate almost unambiguously the
resonance signal from the background given by the conventional
meson-exchange processes (mainly the $t$-channel
$\Delta$--$\Delta$ excitation). It is generally not surprising
that a resonance being short-range in nature is manifested most
pronouncedly in the processes accompanied by large momentum
transfers, where the contributions of peripheral meson-exchange
processes are rather low. This seems the main reason for success
in finding the $\mathcal{D}_{03}$ resonance just in the two-pion
production processes. However, the new polarization measurements
along with the modern partial-wave analysis revealed this
resonance also in $np$ elastic
scattering~\cite{Adl14-el,Adl14-el-2}. Thus, the isoscalar
dibaryon $\mathcal{D}_{03}(2380)$ is quite reliably established up
to date.

Presently there are continuing searches also for other dibaryon
states including those with higher isospins $I=2$ and
$3$~\cite{Clem14}, the supernarrow dibaryons lying below the
pion-production threshold~\cite{Filkov13}, etc. These searches
will definitely be further stimulated by the recent discovery of
pentaquarks by the LHCb Collaboration~\cite{Aaij15}. Pentaquarks
as well as dibaryons had very painful history, so their
experimental finding strongly supports existence of exotic
multiquark states and gives a hope to find more such states in the
near future. However, besides searching for more exotic dibaryon
states in particular processes, it is important to reveal the
interrelations between different dibaryons and different processes
where they can participate, as well as to investigate the role of
dibaryon d.o.f. in short-range $NN$ correlations and in the basic
short-range nuclear force, in general.

It has been suggested recently~\cite{PRC13} that dibaryons can
transform into each other through meson emission and absorption.
In particular, it was shown that the essential role in the decay
of the isoscalar resonance $\mathcal{D}_{03}(2380)$ into $d \pi
\pi$ channel may be played by an intermediate state
$\mathcal{D}_{12}(2150) + \pi$, where $\mathcal{D}_{12}(2150)$ is
an isovector dibaryon with $I(J^P)=1(2^+)$ (also predicted
in~\cite{Dyson64}). This mechanism of the $\mathcal{D}_{03}(2380)$
decay was further incorporated in the rigorous three-body Faddeev
calculations for the $\pi N \Delta$ system~\cite{Gal13}. If the
dibaryon resonances really exist, such transitions between them
seem to be quite natural. In fact, one may present a lot of
examples of similar transitions in the traditional field of baryon
resonances, such as the Roper resonance decay via an intermediate
$\Delta$ isobar: $N^*(1440) \to \Delta + \pi \to N + \pi\pi$.

However, the isovector dibaryons, including
$\mathcal{D}_{12}(2150)$, have not yet become commonly accepted
objects. On the one hand, there are numerous indications for these
dibaryons obtained from both experimental data and several
independent partial-wave analyses (PWA) for the processes $pp
\leftrightarrow pp$, $\pi^+ d \leftrightarrow \pi^+ d$ and $pp
\leftrightarrow \pi^+
d$~\cite{Hoshizaki78,Hoshizaki79,Hoshizaki93,Hoshizaki93-2,Bhandari81,Arndt92,Arndt93,Kravtsov83}.
Besides that, a robust dibaryon pole corresponding to
$\mathcal{D}_{12}(2150)$ was found in the most recent theoretical
calculation~\cite{Gal14} within the framework of Faddeev equations
for the $\pi NN$ system. On the other hand, some previous
amplitude analyses (see, e.g., the
analysis~\cite{Shypit88,Shypit89} of experimental data for the
reaction $pp \to np \pi^+$) did not find a sufficient phase
variation in the dominant $NN \to N\Delta$ partial waves which
could be a signature of dibaryon resonances. The conclusions
against broad dibaryons drawn from the above analysis were later
criticized in a number of works (see,
e.g.,~\cite{Ryskin88,Lee89,Hoshizaki93}), however, were supported
again in~\cite{Anisovich92}. In fact, the dibaryon
$\mathcal{D}_{12}(2150)$, if it exists, lies very near to the
$N\Delta$ threshold and has a width $\Gamma_{\mathcal{D}_{12}}
\simeq 100$--$120$ MeV close to that of the $\Delta$ isobar. So,
one needs very accurate experimental data to distinguish between a
true resonance pole and a threshold cusp in this case.
Unfortunately, as was stated in~\cite{Anisovich92}, the existed
data for $NN \to N\Delta$ amplitudes contained typical
uncertainties of $5$--$10$\%. Moreover, there is a severe
theoretical uncertainty in determining the $NN \to N\Delta$ phase
shift because of the large width of the $\Delta$ isobar. Perhaps
due to these uncertainties, different analyses of experimental
data on reactions $NN \to NN\pi$ have led to controversial
conclusions about existence of a true dibaryon pole near the
$N\Delta$ threshold. Other isovector dibaryons, though lying
higher than $N\Delta$ threshold, have smaller excitation strengths
and larger widths. As a result, the isovector dibaryon resonances
are highly uneasy to identify even in the well studied reactions
$pp \to d \pi^+$ and $pp \to np \pi^+$ where the large momentum
transfers suppress the conventional peripheral processes. The new
high-precision experiments on one-pion production are obviously
needed to shed light on the problem of isovector dibaryons, as was
the case for two-pion production experiments which revealed the
isoscalar resonance
$\mathcal{D}_{03}(2380)$~\cite{Bash09,Adl11,Adl13-iso,Adl12,Adl15}.
Nevertheless, some new important information about both isovector
and isoscalar dibaryon resonances could however still be obtained
from the analysis of different hadronic processes where the same
dibaryons can be excited. Such an analysis is a subject of the
present study.

In the present paper we tried to clarify the question of
intermediate dibaryon contributions in hadronic processes, paying
the most attention to one- and two-pion production in $NN$
collisions. The present work is focused on three main topics:
($i$) revealing the interconnections between different dibaryon
resonances and investigating their possible mutual
transformations; ($ii$) studying the relative role of the same
dibaryons in different hadronic processes; ($iii$) clarifying the
interrelation between the resonance (dibaryon) and background
(meson-exchange) contributions.

The basic motivation of the present study was a general idea that
the processes with large momentum transfers, e.g., $NN \to d \pi$,
$NN \to d \pi\pi$, etc., proceed with a significant probability
through generation of the intermediate resonances, such as
dibaryons, owing to their longer lifetime compared to that of
direct (non-resonance) processes. As the net effect of interaction
is defined by an integral over the interaction time, it should be
easier to transfer a large momentum in a resonance-like process
than in a direct process without time delay.

From the other hand, the $NN$ collisions accompanied by a high
momentum transfer must be very sensitive to the short-range
components of the $NN$ force. Thus, a consistent description of
such processes should apparently take into account the internal
nucleon structure, because one deals here with the inter-nucleon
distances $r_{NN} \lesssim 1$~fm, where the quark cores of two
interacting nucleons are closely overlapped with each other.
However, an explicit account of quark and gluon d.o.f. in
description of hadronic processes like $pp \to d \pi^+$, $pn \to d
\pi^+\pi^-$, etc., would lead to the huge complification of the
whole picture.

At the same time, it was found~\cite{JPG01K,IJMP02K} that the
basic effects of the nucleon quark structure in the $NN$
interaction can be adequately described in terms of dibaryon
rather than quark d.o.f. In such an approach, the $NN$-interaction
$t$-matrix includes several resonance terms of the form
${|\phi_a\rangle\langle\phi_a|}$
$/{(E-M_D^{(a)}+i\Gamma_D^{(a)}/2)}$, where $M_D^{(a)}$ and
$\Gamma_D^{(a)}$ are the mass and width of a dibaryon of the
$a$-th kind, i.e., with a particular set of quantum numbers, and
$|\phi_a\rangle$ is the dibaryon form factor, which represents the
vertex function for the $a$-th resonance decay into $NN$, $NN\pi$,
or $NN\pi\pi$ channels. Such a description does not require an
explicit account of the quark-gluon d.o.f. and is directly related
to the variables of the respective hadronic channel. So, although
the nature of dibaryon resonances is still a subject of
debates~\cite{Bugg14}, their introduction for effective account of
quark d.o.f. at short $NN$ distances appears to be quite
reasonable.

In Sec.~\ref{onepic} the one-pion production process $pp \to d
\pi^+$ is analyzed from the conventional viewpoint. The basic
difficulties in description of this process within the framework
of the conventional meson-exchange approach are demonstrated. Such
a detailed investigation is necessary for clarifying the interplay
between the background (meson-exchange) and resonance (dibaryon)
contributions. In Sec.~\ref{onepid} we explore the contribution of
isovector dibaryons (mainly the $\mathcal{D}_{12}(2150)$) to the
one-pion production processes. By using the realistic parameters
for dibaryon resonances, we obtain a good description for the $pp
\to d \pi^+$ partial and total cross sections. We show further
that the assumed values of dibaryon parameters do not lead to
contradictions in theoretical description of the empirical data
for $pp$ and $\pi^+ d$ elastic scattering. Sec.~\ref{twopi} is
devoted to the analysis of the different $2\pi$-production
processes in $pn$ and $pp$ collisions. The possibility of a
consistent description for one- and two-pion production processes
with inclusion of intermediate dibaryon resonances is
demonstrated. In Sec.~\ref{dibspec} we discuss the possible
quark-cluster structure of dibaryons and its relation to the
observed strong differences between the $2\pi$ production cross
sections in $pn$ and $pp$ collisions in the GeV region. Finally,
in Sec.~\ref{concl} we briefly summarize our conclusions.

\section{Conventional description of the one-pion production
reaction $NN \to d \pi$: problems and solutions} \label{onepic}

The basic one-pion production reaction $NN \to d \pi$ has been the
subject of very numerous experimental and theoretical studies
since 1950ies (see review~\cite{Garc90}). The reaction was treated
within the framework of phenomenological
models~\cite{Grein84,Brack77}, the coupled-channels
approach~\cite{Niskanen78,Niskanen79} and also the Faddeev-type
multiple-scattering approach~\cite{Mizutani81,Lamot87}. Thus it
has long been revealed that the main features of the process at
energies $T_N = 400$--$800$ MeV can be explained by excitation of
an intermediate $N\Delta$ system. The important role is also
played by interference of the $N\Delta$ mechanism with the
one-nucleon-exchange process. From the other hand, the final-state
rescattering contributions were estimated to give no more than
20\% of the total cross section without changing the basic
qualitative features of the reaction~\cite{Grein84}. However, a
number of more sensitive polarization characteristics were not
reproduced within the framework of conventional meson-exchange
models~\cite{Lamot87,Grein84}. So, it was claimed~\cite{Kamo79}
that excitation of the intermediate dibaryon resonances found in
elastic $pp$ scattering~\cite{Makarov82,Auer78,Biegert78} should
be taken into account in the $NN \to d\pi$ process as well.

On the other hand, since one-pion production is accompanied by
rather large momentum transfers ($\Delta p > 350$ MeV), the
contribution of the conventional $N \Delta$ mechanism depends
strongly on the short-range cut-off parameters in the $\pi NN$ and
$\pi N \Delta$ vertices~\cite{Brack77}. Therefore, the proper
choice of these parameters is crucial to determine the real
contribution of the conventional mechanisms and the possible role
of the intermediate dibaryon resonances. To our knowledge, this
important problem, i.e., the relationship between the
contributions of intermediate dibaryons and the values of the
cut-off parameters $\Lambda_{\pi NN}$ and $\Lambda_{\pi N\Delta}$,
has not been paid enough attention in the existing literature.
However, clarification of this issue plays a key role in the
present study. Therefore, after describing the basic formalism for
the reaction $NN \to d \pi$, we consider this problem in detail.

\subsection{Basic formalism}

\begin{figure}[!ht]
\begin{center}
\resizebox{0.8\columnwidth}{!}{\includegraphics{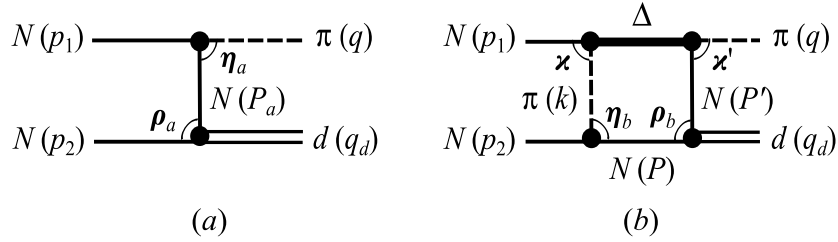}}
\end{center}
\caption{Diagrams illustrating two basic conventional mechanisms
for the reaction $NN \to d \pi$: one-nucleon exchange ($a$) and
intermediate $\Delta$-isobar excitation ($b$). The 4-momenta of
the particles are shown in parentheses, and 3-momenta in pair
center-of-mass systems are denoted by bold face.} \label{fig1}
\end{figure}

Two basic conventional mechanisms of the reaction $NN \to d \pi$,
i.e., one-nucleon exchange\footnote{The one-nucleon-exchange
mechanism of the reaction $NN \to d \pi$ is often referred to in
the literature as an impulse approximation~\cite{Grein84,Brack77}.
However, we prefer to imply under the impulse approximation its
standard meaning, i.e., single scattering in elastic processes.}
and excitation of the intermediate $N\Delta$ system by the
$t$-channel pion exchange are shown in Fig.~\ref{fig1}~$(a)$ and
$(b)$, respectively. Further on, we will refer to these mechanisms
as ONE and $N\Delta$. An excitation of the intermediate $\Delta$
isobar through the $\rho$-meson exchange was also often considered
in the literature~\cite{Brack77}, but such a mechanism contributes
significantly only when choosing very high cut-off parameters in
the meson-baryon form factors. Here, we choose the soft values for
the cut-off parameters\footnote{In the present paper, we assume
$\hbar=c=1$, so the particle masses and momenta are measured in
energy units.} $\Lambda < 1$ GeV (reasons for this are given
below), for which the contribution of the $\rho$-exchange
mechanism is very small.

Relativistic helicity amplitudes corresponding to the diagrams
depicted in Fig.~\ref{fig1} can be written as follows:
\begin{equation*}
\mathcal{M}^{(\rm{ONE})}_{\lambda_1,\lambda_2;\lambda_d} = I_a \,
(2m)^2
\sum\limits_{\lambda'}[\bar{v}(p_2,\lambda_2)G_{dNN}(\lambda_d)u(P',\lambda')]
\end{equation*}
\begin{equation}
\times \frac{1}{P'^2-m^2+i0}[\bar{u}(P',\lambda')F_{\pi
NN}\gamma^{\mu}q_{\mu}\gamma_5 u(p_1,\lambda_1)], \label{eq1}
\end{equation}
\begin{equation*}
\mathcal{M}^{(N\Delta)}_{\lambda_1,\lambda_2;\lambda_d} = I_b
(2m)^2 \sum\limits_{\lambda,\lambda '} \int \frac{i d^4
P}{(2\pi)^4}
\frac{1}{k^2-m_{\pi}^2+i0}[\bar{v}(p_2,\lambda_2)F_{\pi
NN}\gamma^{\mu}k_{\mu}\gamma_5 v(P,\lambda)]
\end{equation*}
\begin{equation}
\times \frac{1}{P^2-m^2+i0}
[\bar{v}(P,\lambda)G_{dNN}(\lambda_d)u(P',\lambda')]\frac{1}{P'^2-m^2+i0}\mathcal{M}^{(\pi
N)}_{\lambda',\lambda_1}, \label{eq2}
\end{equation}
where $\mathcal{M}^{(\pi N)}_{\lambda',\lambda_1}$ is the $\pi
N$-scattering amplitude via an intermediate $\Delta$ isobar:
\begin{equation}
\mathcal{M}^{(\pi N)}_{\lambda',\lambda_1} = - 4mW_{\Delta}
\bar{u}(P',\lambda') \frac{F'_{\pi
N\Delta}q^{\alpha}\mathcal{P}^{(3/2)}_{\alpha\beta}k^{\beta}
F_{\pi N\Delta}}{W_{\Delta}^2 - M_{\Delta}^2 + i
W_{\Delta}\Gamma_{\Delta}(W_{\Delta})}u(p_1,\lambda_1).
\label{eq3}
\end{equation}
The $G_{dNN}$ in Eqs.~(\ref{eq1}) and (\ref{eq2}) stands for the
relativistic deuteron vertex, $I_a$ and $I_b$ are the isospin
coefficients and $\mathcal{P}^{(3/2)}_{\alpha\beta}$ in
Eq.~(\ref{eq3}) denotes the projection operator for the
intermediate $\Delta$. The nucleon spinors are normalized as
$\bar{u}u=-\bar{v}v=1$. The vertex form factors $F_{\pi NN}$ and
$F_{\pi N\Delta}$ will be defined below.

Since not only the reaction amplitudes $\mathcal{M}$ defined in
Eqs.~(\ref{eq1})--(\ref{eq3}), but also each elementary amplitude
(enclosed in square brackets in Eqs.~(\ref{eq1}) and (\ref{eq2}))
are relativistically invariant, it is convenient to calculate each
elementary amplitude in its own c.m.s. Then the resulted
expressions for the amplitudes can be cast into a non-relativistic
form, up to a some energy-dependent factor of relativistic nature.
The explicit form of this factor depends on the specific choice of
the relativistic vertex and often cannot be determined
unambiguously, hence we assume all such factors to be unity.
Neglecting also the small effects of relativistic spin rotations
for the intermediate nucleons, we can write the total amplitude in
terms of nonrelativistic vertices depending on the relative
3-momenta in pairs of particles. Finally, applying the standard
approximation of the spectator nucleon~\cite{Brack77}
\begin{equation}
\label{spec} \int \frac{i d^4 P}{(2\pi)^4}
\frac{2m}{P^2-m^2+i0}\Big|_{P_0=\sqrt{{\bf P}^2+m^2}} \to \int
\frac{d^3 P}{(2\pi)^3}
\end{equation}
and introducing the deuteron wavefunction (d.w.f.)
\begin{equation}
\bar{v}(P)G_{dNN}u(P')\frac{\sqrt{2m}}{P'^2-m^2+i0} \to
-\chi^{\dag}i \sigma_2\Psi^*_{d}(\bm{\rho})\chi,
\end{equation}
one gets the following expressions for the above amplitudes:
\begin{equation}
\mathcal{M}^{(\rm{ONE})}_{\lambda_1,\lambda_2;\lambda_d} = -I_a
(2m)^{3/2} \chi^{\dag}(\lambda_2)i \sigma_2
\Psi^*_{d}(\bm{\rho_a},\lambda_d)F_{\pi
NN}(\eta_a)(\bm{\sigma\eta_a}) \chi(\lambda_1), \label{onem}
\end{equation}
\begin{equation*}
\mathcal{M}^{(N\Delta)}_{\lambda_1,\lambda_2;\lambda_d} = -I_b
(2m)^{1/2} \chi^{\dag}(\lambda_2)i \sigma_2 \int \frac{d^3
P}{(2\pi)^3} \frac{F_{\pi
NN}(\eta_b)(\bm{\sigma\eta_b})}{w_{\pi}^2-m_{\pi}^2+i0}
\end{equation*}
\begin{equation}
\times \Psi^*_{d}(\bm{\rho_b},\lambda_d)
\sqrt{\frac{\Gamma_{\Delta}(\varkappa)\Gamma_{\Delta}(\varkappa')}{\varkappa^3
\varkappa'^3}} \frac{16 \pi W_{\Delta}^2 (\bm{\varkappa\varkappa'}
 + i \frac{\bm{\sigma}}{2}\bm{\varkappa\times\varkappa'})}{W_{\Delta}^2 - M_{\Delta}^2 + i
W_{\Delta}\Gamma_{\Delta}(W_{\Delta})}\chi(\lambda_1),
\label{ndem}
\end{equation}
where $w_{\pi}^2=k^2$ and we used the relation of the
$\Delta$-isobar width to the vertex function $F_{\pi N\Delta}$:
\begin{equation}
\Gamma_{\Delta}(\varkappa) = \frac{\varkappa^3 m}{6\pi
W_{\Delta}}F_{\pi N\Delta}^2(\varkappa). \label{gf}
\end{equation}
To calculate the spin structure of the amplitudes, it is
convenient to write the d.w.f. in the form
\begin{equation}
\label{dwf} \Psi_{d}(\bm{\rho},\lambda_d) = \bm{\sigma}{\bf
E}(\bm{\rho},\lambda_d),
\end{equation}
where we introduced the vector
\begin{equation}
\label{e-vect} {\bf E}(\bm{\rho},\lambda_d) \!=\!
u(\rho)\bm{\varepsilon}(\lambda_d) \! + \!
\frac{w(\rho)}{\sqrt{2}} \!\! \left(\bm{\varepsilon}(\lambda_d)
\!-\!
\frac{3\bm{\rho}(\bm{\rho\varepsilon}(\lambda_d))}{\rho^2}\right).
\end{equation}
Here, $\bm{\varepsilon}(\lambda_d)$ is the standard deuteron
polarization vector, $u$ and $w$ are the $S$- and $D$-wave
components of the d.w.f. normalized as $\int d^3 \rho \left(u^2 +
w^2\right)/ {(2\pi)^3} = 1$.

Although the vertices in Eqs.~(\ref{onem}) and (\ref{ndem}) are
calculated non-relativistically, we still employ relativistic
kinematics in calculations of the relative momenta, according to
the minimal relativity principle. In fact, comparison of the
results of non-relativistic~\cite{Brack77} and fully
relativistic~\cite{Grein84} calculations for the ONE and $N\Delta$
mechanisms shows that the account of relativistic effects as well
as the deviation from the nucleon-spectator approximation give a
correction of no more than 10--15\%. Since the description of the
$NN \to d \pi$ reaction in terms of two basic mechanisms only is
initially approximate, the fully relativistic description of these
mechanisms, requiring much more elaborated calculations, seems
impractical at this stage. Furthermore, since the relativistic
factors which we neglected here would increase the cross sections
by 10--15\% and the rescattering corrections would, on the
contrary, decrease them by $\simeq 20$\%~\cite{Grein84}, these two
types of corrections would considerably cancel each other.

For definiteness, the reaction $pp \to d \pi^+$ will be considered
further. Then the isospin coefficients are $I_a = \sqrt{2}$ and
$I_b = 4\sqrt{2}/3$. The helicity amplitudes must be
antisymmetrized over the initial protons. Then they take the
form\footnote{The factor $1/\sqrt{2}$ appearing in Eq.~(A8) of
Ref.~\cite{Grein84} as well as the same factor for the
$d$--$n$--$p$ isospin vertex are included here in the d.w.f.
normalization.}~\cite{Grein84}
\begin{equation}
\mathcal{M}^{(s)}_{\lambda_1,\lambda_2;\lambda_d}(\theta) =
\mathcal{M}_{\lambda_1,\lambda_2;\lambda_d}(\theta) +
(-1)^{\lambda_d}\mathcal{M}_{\lambda_2,\lambda_1;\lambda_d}(\pi-\theta).
\end{equation}
Overall, there are 6 independent helicity amplitudes in the
reaction $pp \to d \pi^+$:
\begin{equation*}
\Phi_1 = \mathcal{M}^{(s)}_{\frac{1}{2},\frac{1}{2};1}, \quad
\Phi_2 = \mathcal{M}^{(s)}_{\frac{1}{2},\frac{1}{2};0}, \quad
\Phi_3 = \mathcal{M}^{(s)}_{\frac{1}{2},\frac{1}{2};-1},
\end{equation*}
\begin{equation}
\Phi_4 = \mathcal{M}^{(s)}_{\frac{1}{2},-\frac{1}{2};1}, \quad
\Phi_5 = \mathcal{M}^{(s)}_{\frac{1}{2},-\frac{1}{2};0}, \quad
\Phi_6 = \mathcal{M}^{(s)}_{\frac{1}{2},-\frac{1}{2};-1}.
\end{equation}
The total cross section is expressed through the above six
amplitudes as follows:
\begin{equation}
\sigma(pp \to d \pi^+) = \frac{1}{64\pi
s}\frac{q}{p}\int\limits_{-1}^{1}\sum\limits_{i=1}^{6}\left|\Phi_i(x)\right|^2
dx,
\end{equation}
where $p$ and $q$ are the moduli of the proton and the pion c.m.s.
momenta, respectively, and $x = \rm{cos}(\theta)$.

For comparison of the theoretical results with the PWA data and
for studying the contributions of the intermediate dibaryon
resonances, it is convenient to deal with the partial-wave
amplitudes, which are expressed through the helicity ones via the
standard formulas given by Jacob and Wick~\cite{Jacob59}. We
display here the explicit formulas for the amplitudes in two
dominant partial waves $^{2S+1}L_JL^{\pi d} = {}^1D_2P$ and
$^3F_3D$ only:\footnote{$S$ and $L$ are related to the $NN$
system.}
\begin{equation}
A(^1D_2P) = \frac{1}{2}\sqrt{\frac{3}{5}}\left(\Phi_1^{(2)} +
\Phi_3^{(2)}\right) + \frac{1}{\sqrt{5}}\Phi_2^{(2)}, \label{abg}
\end{equation}
\begin{equation}
A(^3F_3D) = -\frac{2}{\sqrt{7}}\Phi_4^{(3)} -
\frac{1}{2}\sqrt{\frac{6}{7}} \Phi_5^{(3)},
\end{equation}
where
\begin{equation}
\Phi_i^{(J)} =
\int\limits_{-1}^{1}d^{(J)}_{\lambda_1-\lambda_2,-\lambda_d}(x)\Phi_i(x)dx.
\end{equation}
The respective partial cross sections are
\begin{equation}
\sigma(^{2S+1}L_JL^{\pi d}) = \frac{(2J+1)}{64\pi
s}\frac{q}{p}\left|A(^{2S^{pp}+1}L^{pp}_JL^{\pi d})\right|^2.
\end{equation}

\subsection{Parametrization of the vertex form factors:
the cut-off problem} The main issue in the calculation of the
amplitudes for the conventional processes, such as $N\Delta$
mechanism shown in Fig.~\ref{fig1}~$(b)$, is the parametrization
of the meson-baryon vertex functions, in our case, the $F_{\pi
NN}$ and $F_{\pi N\Delta}$, especially in the short-range (or
high-momentum) region. In fact, the exact form of these vertex
functions and the true values for the short-range cut-off
parameters $\Lambda_{\pi NN}$ and $\Lambda_{\pi N\Delta}$ are
still unknown, despite the very numerous works dedicated to this
problem (see, e.g.,~\cite{Koepf96} and references therein).
However, results of the different quark-model-based calculations
agree, in general, that these parameters should be essentially
soft ($\Lambda = 0.4$--$0.9$~GeV)~\cite{Koepf96}.

In the present study, we have chosen the most simple vertex
parametrization which follows directly from the basic principles
of non-relativistic quantum mechanics combined with a minimal
relativity principle. The advantages of such a choice are
demonstrated below.

In the $\pi N$ c.m.s., the vertex functions $F_{\pi NN}$ and
$F_{\pi N\Delta}$ depend on the relative momentum of the pion and
the nucleon. In its turn, the modulus of the relative momentum of
two particles $b$ and $c$ produced in the decay of a particle $a$
is a relativistically invariant quantity depending on invariant
masses of all three particles:
\begin{equation}
p_{bc}^2 = \frac{\left(w_a^2 - w_b^2 - w_c^2\right)^2 - 4 w_b^2
w_c^2}{4 w_a^2}.
\end{equation}
Then, writing the vertex form factor as a functions of $p_{bc}$
makes it possible to describe \emph{the real and virtual particles
in a unified manner}.

When choosing a simple monopole parametrization for the above
vertex functions, one has:
\begin{equation}
\label{fpinn} F_{\pi NN}(p,\tilde{\Lambda}) =
\frac{f}{m_{\pi}}\frac{p_0^2 + \tilde{\Lambda}^2}{p^2 +
\tilde{\Lambda}^2},
\end{equation}
\begin{equation}
\label{fpind} F_{\pi N \Delta}(p,\tilde{\Lambda}_*) =
\frac{f_*}{m_{\pi}}\frac{p_{0}^2 + \tilde{\Lambda}_*^2}{p^2 +
\tilde{\Lambda}_*^2},
\end{equation}
where $p^2$ is a modulo squared of the $\pi$--$N$ relative
momentum (i.e., the pion momentum in the $\pi N$ c.m.s.) and
$p_0^2$ corresponds to the situation when all three particles are
real, i.e., located on their mass shells. Then one gets the
standard expression for the $\Delta \to \pi N$ decay width (see
Eq.~(\ref{gf})):
\begin{equation}
\label{gdel} \Gamma_{\Delta}(p) = \Gamma_{\Delta}
\left(\frac{M_{\Delta}}{W_{\Delta}}\right)
\left(\frac{p}{p_0}\right)^3 \left(\frac{p_0^2 +
\tilde{\Lambda}_*^2}{p^2 + \tilde{\Lambda}_*^2}\right)^2.
\end{equation}
The coupling constants in Eqs.~(\ref{fpinn})--(\ref{fpind}) have
been taken to be $f = 0.97$ and $f_* = 2.17$. In this case, one
has $f^2/4\pi=0.075$, and the above value for $f_*$ was derived
from the total width of the $\Delta$ isobar $\Gamma_{\Delta} =
117$ MeV as given by the Particle Data Group~\cite{PDG14}.

In case when \emph{only pion} is off the mass shell,
Eqs.~(\ref{fpinn})--(\ref{fpind}) are reduced to the standard
monopole form factors, depending on the pion invariant mass
$w_{\pi}$ only (up to small terms proportional to $w_{\pi}^4$):
\begin{equation}
F_{\pi NN}(w_{\pi};w_N = m,w_N = m) \simeq
\frac{f}{m_{\pi}}\frac{m_{\pi}^2 - {\Lambda}^2}{w_{\pi}^2 -
{\Lambda}^2},
\end{equation}
\begin{equation}
F_{\pi N \Delta}(w_{\pi};w_N = m,w_{\Delta} = M_{\Delta}) \simeq
\frac{f_*}{m_{\pi}}\frac{m_{\pi}^2 - {\Lambda_*}^2}{w_{\pi}^2 -
{\Lambda_*}^2},
\end{equation}
where the cut-off parameters are related to the initial ones by
\begin{equation}
{\Lambda}^2 \simeq \tilde{\Lambda}^2, \quad {\Lambda_*}^2 \simeq
\left(\tilde{\Lambda}_*^2 + \left(\frac{M_{\Delta}^2 -
m^2}{2M_{\Delta}}\right)^2\right)\Bigg/\left(\frac{M_{\Delta}^2 +
m^2}{2M_{\Delta}^2}\right). \label{la12}
\end{equation}

It should be noted here that a different parametrization for the
phenomenological vertices of the type $F_{a \to bc}$ is often used
in the literature. In this commonly used parametrization, the
total vertex function is represented as a product of three
independent functions, each depending on the one invariant mass
only (see, e.g.,~\cite{Schutz96}). This form of the vertices
contains at least three independent parameters, some of which
cannot be found from experimental data. Therefore, such a
parametrization does not allow to establish a direct
interconnection between the different processes involving the same
particles \emph{on} and \emph{off} the mass shell. On the other
hand, the vertex parametrization of the form $F(p_{bc},\Lambda)$
with a single cut-off parameter $\Lambda$, used in the present
work, is consistent with the basic principles of quantum mechanics
and admits a straightforward off-shell continuation. The parameter
$\Lambda$ in such a case can in general be found directly from
experimental data.

\begin{figure}[!ht]
\begin{center}
\resizebox{0.6\columnwidth}{!}{\includegraphics{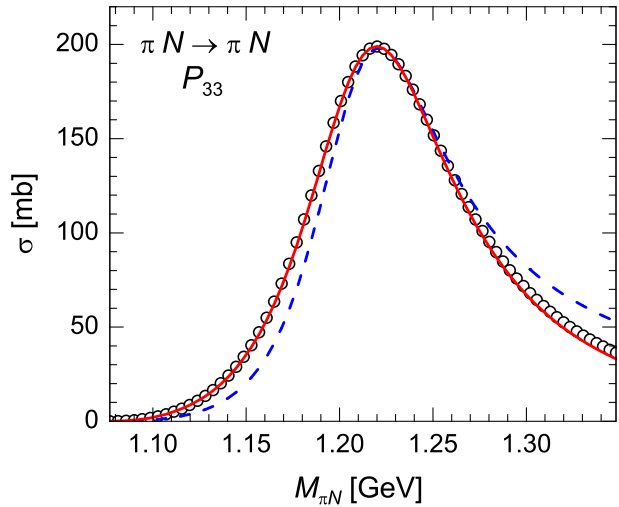}}
\end{center}
\caption{(Color online) The cross section of $\pi N$ elastic
scattering in the $P_{33}$ partial wave. Solid and dashed lines
show the calculations in the isobar model with the $\pi N \Delta$
vertex in the form (\ref{fpind}) and the cut-off parameters
$\tilde{\Lambda}_* = 0.3$ and $0.55$~GeV, respectively. Solid
circles correspond to the PWA data (SAID, solution
WI08~\cite{SAID}).} \label{fig2}
\end{figure}

Thus, the parameter $\tilde{\Lambda}_*$ in the $\pi N \Delta$
vertex can be found from empirical data on $\pi N$ elastic
scattering. Fig.~\ref{fig2} shows the PWA (SAID) data~\cite{SAID}
for the $\pi N$-scattering cross section in the $P_{33}$ partial
wave and the results of calculations in the isobar model with a
vertex form factor~(\ref{fpind}) for two values of the parameter
$\tilde{\Lambda}_*$. We found that the best agreement between the
theoretical calculation and the empirical data in a wide energy
range is obtained by choosing the value $\tilde{\Lambda}_* = 0.3$
GeV.\footnote{We note in passing that a similar value
$\tilde{\Lambda}_* \simeq 0.36/ \sqrt{2} \simeq 0.26$ GeV (where
we used the relation of the monopole cut-off parameter to the
dipole one) was taken in Ref.~\cite{Lee84} to describe the
$NN$-scattering phase shifts up to the energies $T_N = 2$ GeV
consistently with $\pi N$ elastic scattering.} Then, using
Eq.~(\ref{la12}), we obtain the respective monopole parameter
$\Lambda_* = 0.44$ GeV, which is indeed very soft.

It was argued in a number of theoretical works that the cut-off
parameter value in the $\pi N \Delta$ vertex (in the monopole
form) should be substantially (100--300 MeV) less than that in the
$\pi NN$ vertex (see, e.g.,~\cite{Koepf96,Gari95,Kondr70}). In the
present study, we have taken the value $\Lambda \simeq
\tilde{\Lambda} = 0.7$ GeV, which was used in a number of previous
calculations of reactions such as $NN \to d
\pi$~\cite{Grein84,Uzikov88}. This value of $\Lambda$ is
consistent with the predictions of the lattice-QCD
calculations~\cite{Liu99,Erkol09} (see also
Ref.~\cite{Plessas09}). Thus, the monopole fits for the results
obtained in~\cite{Liu99} (lattice QCD with extrapolation to the
physical pion mass) and~\cite{Erkol09} (extrapolation to the
chiral limit) give $\Lambda = 0.75$ and $0.61$ GeV, respectively.
We emphasize here that the similar values $\Lambda = 0.65$--$0.7$
GeV were obtained in the fit of $NN$-scattering phase shifts and
the deuteron properties within the dibaryon model for $NN$
interaction~\cite{JPG01K}. One should also note that relativistic
quark models predict an even softer cut-off for the ${\pi NN}$
vertex function~\cite{Plessas09}. Unfortunately for the $\pi N
\Delta$ form factor, we presently have no lattice-QCD predictions
at the physical pion mass (or the respective extrapolation), and
the available results at $m_{\pi} \simeq$ 300 MeV~\cite{Alex11}
give too high cut-off parameters for both $\pi N N$ and $\pi N
\Delta$ vertices. Therefore, one is forced to use phenomenological
parametrizations for $F_{\pi N \Delta}$, like the one used in this
work, trying to relate the parameters to experimental data
wherever possible.

So, for the ratio of the cut-off parameters in the vertices
$F_{\pi NN}$ and $F_{\pi N \Delta}$, we obtained the value
${\Lambda_*}/{\Lambda} \simeq 0.6$. Note that the same value was
derived in~\cite{Huber94} from comparison of the relativistic
meson-exchange model calculations with experimental data for the
process $NN \to NN\pi$.

It should be stressed here once again that the parametrization for
the vertex functions $F_{\pi NN}$ and $F_{\pi N \Delta}$ which we
adopt in the present study describes the real and virtual
particles in a unified manner. Hence, it can be used for
consistent description of different processes involving on- and
off-shell pions, i.e., $\pi N \to \pi N$, $NN \to \pi d$, elastic
$NN$ scattering, etc., with the same realistic (soft) cut-off
parameters in the meson-baryon vertices. It does not require
introducing any additional parameters to account for the pion
virtuality. Although this choice of the vertex parametrization is
not unique, it seems to be the simplest and most natural one.

It also should be stressed that the cut-off parameters used here
are much softer than those traditionally used in the realistic
$NN$-potential models. For example, in the Bonn
model~\cite{Machl87}, the minimal values, which still allow a good
description of $NN$-scattering phase shifts up to $T_N = 350$ MeV,
are $\Lambda \simeq \Lambda_* \simeq 1.3$ GeV (in the upgraded
CD-Bonn model~\cite{Machl01} these values are even higher). Such
very high cut-off parameters apparently lead to increased
meson-exchange contributions at small inter-nucleon distances. In
many cases, however, the artificial strengthening of the
meson-exchange processes can mimic somehow the contributions of
short-range QCD-based mechanisms which involve the quark-meson
structure of interacting nucleons. So, in this way, the
$t$-channel meson-exchange mechanisms with artificially enhanced
cut-off parameters can really give the correct behaviour of some
observables. For example, as was shown in Ref.~\cite{Ericson82},
an accurate description of the basic deuteron properties can be
obtained in the simple meson-exchange model, which takes into
account the one-pion exchange only, without any cut-off, i.e.,
with $\Lambda = \infty$. One can suggest this continuity between
the peripheral meson-exchange and short-range QCD-based mechanisms
to be a manifestation of a fundamental quark/hadron continuity
principle.

On the other hand, from the fact that the vertex function $F_{\pi
N \Delta}$ in $\pi N$ scattering should have $\Lambda_* \simeq
0.4$ GeV, while for the description of reactions like $NN \to d
\pi$ one should take $\Lambda_* \simeq 0.6$ GeV (see below and
also Ref.~\cite{Uzikov88}), and at the same time the correct
description of the deuteron properties and $S$-wave
$NN$-scattering requires $\Lambda_* \simeq 1.3$
GeV~\cite{Machl87}, it follows that the phenomenological approach
based on \emph{ad hoc} fitting the short-range cut-off parameters
in the meson-baryon vertices to describe a specific process is not
quite consistent, and probably contains some internal
contradictions tightly related to the contributions of quark
d.o.f. (see Ref.~\cite{Holinde92} for the detailed discussion on
this issue). Instead, one could use the \emph{universal}
(essentially soft) cut-off parameters in the meson-baryon vertices
to describe \emph{different processes in a unified manner}. Then
the deviations from experimental data, which would inevitably
arise in this situation, might be regarded as indications of some
short-range QCD-based mechanisms, not taken into account in the
conventional meson-exchange approach. In this case, the stronger
the observed discrepancies are and, accordingly, the larger
cut-off parameters are needed to describe the experimental data,
the stronger the ``hidden'' quark d.o.f. manifest themselves in
the process in question. We will return to these ideas in
Sec.~\ref{onepid}, where the contributions of intermediate
six-quark objects (dibaryons) will be considered.

As was shown above, the parametrization of the vertices in the
form (\ref{fpinn})--(\ref{fpind}) allows us to take into account
the effects of \emph{any} of the three particles going off the
mass shell. The most noticeable effect due to presence of the
\emph{off-shell nucleons} is seen in the ONE process, where the
nucleon after pion emission is strongly off-shell. Introducing the
form factor (\ref{fpinn}) at $(w_{N'}; w_{\pi} = m_{\pi}, w_N =
m)$ in the vertex $F_{\pi NN'}$ ($N'$ being the nucleon after pion
emission), we found that the ONE contribution is reduced by
$\simeq 30\%$ in comparison with the use of a constant $\pi NN'$
form factor. It should be noted that just the same effect was
obtained in~\cite{Grein84}, where the vertex $F_{\pi NN'}$ with
the off-shell nucleon has been derived from the dispersion
relations. This coincidence provides an additional argument in
favor of the vertex parametrization employed in the present work.
We also got a reduction of the $N\Delta$ mechanism (taken in the
nucleon-spectator approximation) due to the nucleon $N'$ going off
the mass shell, but this effect turned out to be less significant
than in case of the ONE mechanism, and amounted to $10\%$ only.

\subsection{Results and discussion}

\begin{figure}[!ht]
\begin{center}
\resizebox{1.0\columnwidth}{!}{\includegraphics{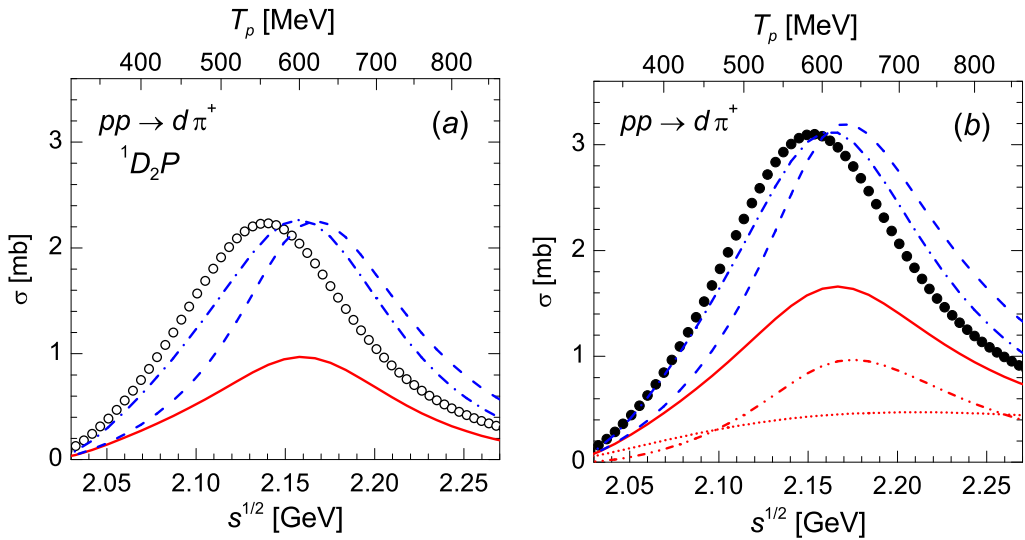}}
\end{center}
\caption{(Color online) ($a$) The partial $pp \to d \pi^+$ cross
section in the dominant $^1D_2P$ partial wave calculated within
the ONE + $N\Delta$ model. Solid line --- calculation with the
vertex form factors (\ref{fpinn})--(\ref{fpind}) and parameter
values $\tilde{\Lambda} = 0.7$ and $\tilde{\Lambda}_* = 0.3$ GeV,
(corresponding to the monopole parameters $\Lambda = 0.7$ and
$\Lambda_* = 0.44$ GeV --- see Eq.~(\ref{la12})). Dashed line
--- the same vertex parametrization used but with $\tilde{\Lambda}_* = 0.55$ GeV
(monopole $\Lambda_* = 0.68$ GeV). Dash-dotted line ---
calculation with form factors (\ref{fpinn-2})--(\ref{fpind-2}),
${\Lambda} = 0.7$ and $\Lambda_* = 0.6$ GeV. The open circles
correspond to the PWA data (SAID, solution C500~\cite{SAID,Oh97}).
($b$) The same as ($a$) but for the total $pp \to d \pi^+$ cross
section. The individual contributions of the ONE and $N\Delta$
mechanisms are shown by dotted and dash-dot-dotted lines. The PWA
results (coinciding with experimental data) are shown by filled
circles.} \label{fig3}
\end{figure}

We calculated the cross sections for the one-pion production
reaction $pp \to d\pi^+$ in the energy range $\sqrt {s} =
2.03$--$2.27$ GeV ($T_p \simeq 320$--$860$ MeV) using the above
formalism. The results for the partial cross section in the
dominant partial wave $^1D_2P$ and for the total cross section are
shown in Fig.~\ref{fig3}~($a$) and ($b$), respectively. As
``experimental'' data for comparison with theoretical
calculations, we took the results of PWA (SAID, solution
C500~\cite{Oh97}) initially obtained for the inverse reaction
$\pi^+ d \to pp$. The cross sections of the two reactions are
related as
\begin{equation}
\sigma(pp \to d \pi^+) =
\frac{3}{2}\left(\frac{q}{p}\right)^2\sigma(\pi^+ d \to pp).
\end{equation}
The advantage of the chosen PWA solution (C500) is that it was
found in a combined analysis of the three interrelated processes
$\pi^+ d \to pp$, $pp \to pp$ and $\pi^+ d \to \pi^+ d$. The
results of this PWA solution for the reaction $\pi^+ d \to pp$ in
the dominant partial waves $^1D_2P$, $^3F_3D$, etc., are in good
agreement with the older PWA results~\cite{Hiroshige84,Bugg88}.

Because of the high transferred momenta in the one-pion production
process ($\Delta p > 350$ MeV), the theoretical predictions can be
expected to be sensitive to the model of the deuteron wave
function (d.w.f.) used in calculations. To clarify this issue, we
considered two models for the d.w.f. --- the one derived from the
CD-Bonn potential model~\cite{Machl01} and the one obtained from
the dibaryon model for $NN$ interaction~\cite{IJMP02K}. Both wave
functions describe the observable deuteron properties well, but
have different behaviour in the high-momentum region. Our study
has shown that, although the ONE mechanism is indeed very
sensitive to the choice of the d.w.f., the effect of using
different d.w.f. models in the summed contribution of the ONE + $N
\Delta$ mechanisms to the partial ($^1D_2P$) and total cross
sections does not exceed 10\%. So, we present here the results for
the dibaryon d.w.f. only.

We found that the conventional ONE + $N \Delta$ mechanisms with
the meson-baryon vertices parameterized in the form
(\ref{fpinn})--(\ref{fpind}) using the parameters $\tilde{\Lambda}
= 0.7$ and $\tilde{\Lambda}_* = 0.3$ GeV (corresponding to the
monopole parameters ${\Lambda} = 0.7$ and ${\Lambda}_* = 0.44$
GeV) give about half the experimental cross section in the
$^1D_2P$ partial wave and also about half the total cross section
near their maximal values (at $\sqrt{s} \simeq 2.14$--$2.16$ GeV),
with a theoretical peak shifted by about 20 MeV to the right
relatively to its experimental position. It is worth noting that
quite similar results were obtained previously in the
works~\cite{Kamo79,Kamada97}.

On the other hand, due to the strong sensitivity of the results to
the cut-off parameters in vertices $F_{\pi NN}$ and $F_{\pi N
\Delta}$, enhancing these parameter values may lead to a
significant increase in the theoretical cross sections. To
demonstrate the importance of this observation, we examined two
ways of the parameter variation. First, we increased the value of
the parameter $\tilde{\Lambda}_*$ in the $\pi N \Delta$ vertex
from $0.3$ to $0.55$ GeV. In this case, we were able to reproduce
approximately the shape of both partial and total cross sections,
however, with a significant shift of the resonance peak (see
Fig.~\ref{fig3}). On the other hand, as we demonstrated above, the
value $\tilde{\Lambda}_* = 0.55$ GeV is no longer appropriate to
describe the empirical data on the elastic $\pi N$ scattering
beyond the resonance peak (see Fig.~\ref{fig2}).

The second way often used in the literature is changing the vertex
parametrization itself, so that the degree of virtuality for each
of the three particles is governed by its own cut-off parameter
independent from other particles. In this case, the monopole form
factors for the $\pi NN$ and $\pi N \Delta$ vertices with
off-shell pions (and an off-shell $\Delta$) would be written as
follows:
\begin{equation}
\label{fpinn-2} F_{\pi N N}^{(2)}(w_{\pi}) =
\frac{f}{m_{\pi}}\frac{m_{\pi}^2 - {\Lambda}^2}{w_{\pi}^2 -
{\Lambda}^2},
\end{equation}
\begin{equation}
\label{fpind-2} F_{\pi N \Delta}^{(2)}(W_{\Delta};w_{\pi}) =
\frac{f_*}{m_{\pi}}\frac{{\varkappa}_0^2 +
\tilde{\Lambda}_*^2}{{\varkappa}_{\rm on}^2 +
\tilde{\Lambda}_*^2}\frac{m_{\pi}^2 - {\Lambda}_*^2}{w_{\pi}^2 -
{\Lambda}_*^2},
\end{equation}
where ${\varkappa}_{\rm on}$ is the magnitude of the on-shell
$\pi$--$N$ relative momentum, i.e., at $w_{N}=m$ and
$w_{\pi}=m_{\pi}$, thus dependent on $W_{\Delta}$ only. In such a
parametrization, the parameter $\tilde{\Lambda}_*$ can still be
chosen as to describe the $\pi N$ elastic scattering and thus
should be equal to $0.3$ GeV. The pion virtuality is, however,
controlled by an additional parameter ${\Lambda}_*$, which cannot
be determined from experimental data and thus is fitted to a
particular process (in our case, $pp \to d \pi^+$). So, this way
of vertex parametrization is fully \emph{ad hoc}.

The results of calculations for the partial and total cross
sections within the ONE + $N\Delta$ model using the vertex
parametrization (\ref{fpinn-2})--(\ref{fpind-2}) are also shown in
Fig.~\ref{fig3}. It turns out that theoretical calculations are
approximately consistent in magnitude with the empirical data,
when using $\Lambda_* = 0.6$ GeV.

\begin{figure}[!ht]
\begin{center}
\resizebox{0.6\columnwidth}{!}{\includegraphics{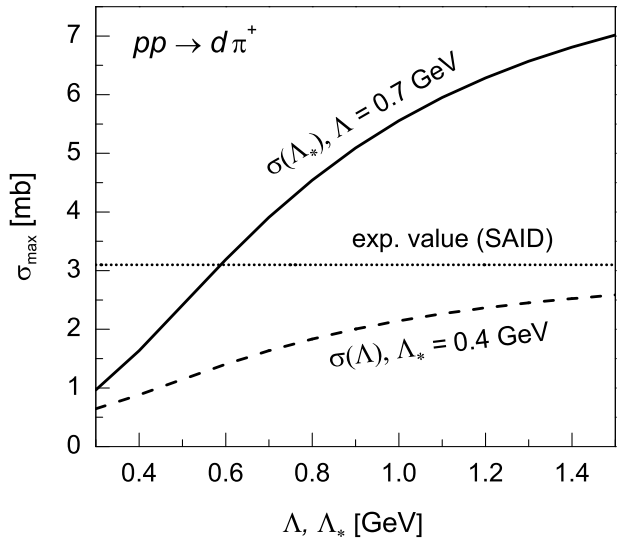}}
\end{center}
\caption{Dependence of the theoretically calculated total cross
section for the reaction $pp \to d \pi^+$ at the peak ($\sqrt{s} =
2.16$ GeV) on the cut-off parameters in the vertices $F_{\pi NN}$
($\Lambda$) and $F_{\pi N \Delta}$ ($\Lambda_*$) (see
Eqs.~(\ref{fpinn-2})--(\ref{fpind-2})) is shown by dashed and
solid lines, respectively. The empirical peak cross section (at
$\sqrt{s} = 2.15$ GeV) according to PWA (SAID) data is shown by
dotted line.} \label{fig4}
\end{figure}

The dependence of the theoretically calculated value of the peak
total cross section (at $\sqrt {s} = 2.16$ GeV) upon the cut-off
parameters in vertices (parameterized in the form
(\ref{fpinn-2})--(\ref{fpind-2})) is shown in Fig.~\ref{fig4}. In
particular, the dependence of the peak cross section on the
parameter $\Lambda$ in the vertex $F_{\pi NN}$ at a fixed value of
$\Lambda_* = 0.4$~GeV in the vertex $F_{\pi N \Delta}$ and the
dependence on the parameter $\Lambda_*$ at a fixed value of
$\Lambda = 0.7$~GeV are presented. For comparison, the
experimental peak cross section value (coincident with the PWA
result) is also shown. It is seen from Fig.~\ref{fig4}, that the
theoretical cross section depends strongly on the cut-off
parameters in vertices, especially on the parameter $\Lambda_*$ in
the $\pi N \Delta$ vertex. Thus, when $\Lambda_*$ is increased by
$50$\%, i.e., from $0.4$ to $0.6$ GeV, the cross section increases
by two times.

So, when enhancing the cut-off parameters in vertices, one is able
to approximately reproduce the experimental height of the cross
section. However, as is seen from Fig.~\ref{fig3}, independently
on the vertex parametrization, the cross section peak remains
shifted to the right relatively to its experimental position. This
energy shift is particularly noticeable for the partial cross
section in the $^1D_2P$ wave, where it amounts to about 20--30
MeV. This result might be considered as an indication of the
contribution of some additional mechanism in this process.

In particular, adding some $N\Delta$ attraction can shift the peak
position of the calculated $pp \to d \pi^+$ cross section
downwards (see, e.g.,~\cite{Niskanen95}). However, in view of the
above problems with meson-baryon form factors and also the large
width of the $\Delta$, the nature of this attraction is not clear
enough. We argue that generation of an intermediate dibaryon
resonance may provide the basic attraction in the $N\Delta$ system
(in addition to the peripheral pion exchange), similarly to that
found in the $NN$ system~\cite{JPG01K}. So, excitation of the
intermediate dibaryon $\mathcal{D}_{12}(2150)$ in the $^1D_2P$
partial wave seems to be a likely candidate for the missing
mechanism. Indeed, according to the numerous
predictions~\cite{Dyson64,Bhandari81,Gal14}, the mass of this
dibaryon lies about 10--30 MeV below the $N\Delta$ threshold.
Besides that, although the discrepancies for the total cross
section could in principle be eliminated by conventional
mechanisms, significant disagreement with experimental data
remains in the more sensitive spin-dependent observables, even
after taking all rescattering corrections (and also relativistic
effects) into account. Particularly strong disagreement was
revealed in the tensor analyzing powers~\cite{Grein84,Lamot87}. We
postpone the detailed investigation of the observables (including
spin-dependent ones) in the reaction $pp \to d \pi^+$ for our next
paper. Here, we would just like to show that a consistent
description of one-pion production by the conventional
meson-exchange mechanisms encounters serious difficulties,
including those which cannot be eliminated by fitting the cut-off
parameters in the vertex form factors.

Thus, in this section we have demonstrated some problems faced by
the conventional meson-exchange models in description of hadronic
processes with high momentum transfers, in particular, of one-pion
production. The main difficulty lies in the strong sensitivity to
the short-range cut-off parameters in the vertices of meson
emission and absorption. It is rather obvious that, until these
parameters are determined accurately from the fundamental theory,
one will not be able to reveal the real degree of discrepancy
between the traditional meson-exchange model calculations and
experimental data. Nevertheless, it appears to be a general trend
that describing the processes involving two nucleons requires
higher cut-off parameters in the meson-baryon vertices than the
processes involving just one nucleon. Hence, instead of increasing
the cut-off parameters \emph{ad hoc} to describe the one-pion
production (at the cost of consistency with other processes), one
can try to find the missing contributions by including the
resonance mechanisms based on the assumption of the intermediate
dibaryon formation.

\section{Inclusion of intermediate (isovector) dibaryons in one-pion production and elastic scattering}
\label{onepid}

\subsection{Reaction $pp \to d \pi^+$ with intermediate dibaryons}

\begin{figure}[!ht]
\begin{center}
\resizebox{0.4\columnwidth}{!}{\includegraphics{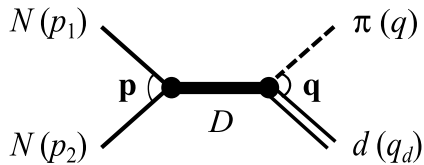}}
\end{center}
\caption{Diagram illustrating the excitation of an intermediate
dibaryon resonance in the reaction $NN \to d \pi$.} \label{fig5}
\end{figure}

Let us now consider how the partial cross section of the reaction
$pp \to d\pi^+$ in the dominant $^1D_2P$ wave changes, if one adds
to the background amplitude determined by the ONE + $N\Delta$
mechanisms (see Eq.~(\ref{abg})) the resonance amplitude
corresponding to excitation of the intermediate dibaryon
$\mathcal{D}_{12}(2150)$. A diagram illustrating such a resonance
mechanism is shown in Fig.~\ref{fig5}. The respective partial-wave
amplitude has the form
\begin{equation}
A^{(D)}(^1D_2P) = -\frac{8\pi s}{\sqrt{pq}}\frac{\sqrt{2
\Gamma_{D_{12} \to pp}(s)\,\Gamma_{D_{12} \to \pi
d}(s)}}{s-M_{D_{12}}^2+i\sqrt{s}\Gamma_{D_{12}}(s)}.
\label{eq-dib}
\end{equation}
The factor 2 before the partial width $\Gamma_{D_{12} \to pp}$ is
introduced to account for the identical particles in the initial
state.

To calculate the contribution of an intermediate dibaryon to a
particular process, one has to fix the dibaryon parameters
somehow. It should be noted however that the parameters of
dibaryon resonances and especially their partial widths are
presently known with large uncertainties. The vast majority of
phenomenological studies of dibaryon contributions to hadronic
processes carried out in 1980ies (see,
e.g.,~\cite{Kamo79,Kanai79,Barannik86}) included \emph{ad hoc}
fitting the parameters of dibaryon resonances to a particular
process in question. For instance, in the work~\cite{Kamo79} which
considered dibaryon contributions to the process $pp \to d \pi^+$,
the parameters of six hypothetical dibaryons were simultaneously
fitted to describe the experimental data. So, it was highly uneasy
to draw some reliable conclusions about the real contribution of
intermediate dibaryons to this process. Contrary to this, we took
reasonable values for the basic parameters of dibaryon resonances
from existing literature or from the clear physical
considerations, and then tested the sensitivity of the obtained
results to the parameter variation. Presently, plausible estimates
can be found in the literature at least for the most reliably
established dibaryons $\mathcal{D}_{12}(2150)$ and
$\mathcal{D}_{03}(2380)$.

For the dibaryon $\mathcal{D}_{12}$, we first fixed (up to $\pm
10$~MeV) its mass and total width to be $M_{D_{12}} = 2.15$ GeV
and $\Gamma_{D_{12}} = 110$~MeV. This choice was based on the PWA
results~\cite{Hoshizaki93,Hoshizaki93-2} and also on the results
of the recent Faddeev calculations for the $\pi NN$
system~\cite{Gal14}. For parametrization of the energy dependence
of the total resonance width, we took the form $\Gamma_{D_{12}}(s)
= \Gamma_{D_{12}\to \pi d}(s)/R_{\pi d}$, where $R_{\pi d} =
\Gamma_{D_{12} \to \pi d} / \Gamma_{D_{12}}$ is the branching
ratio for the $\mathcal{D}_{12} \to \pi d$ decay mode at the
resonance point. In other words, we assumed that the total width
of the $\mathcal{D}_{12}$ dibaryon is proportional to its partial
decay width into the $\pi d$ channel. This assumption was based on
the fact that the decay $\mathcal{D}_{12} \to \pi NN$ has the same
threshold behaviour and the same dynamical mechanism in its origin
as the decay $\mathcal{D}_{12} \to \pi d$, and the partial width
$\Gamma_{D_{12} \to NN}$, according to a number of
estimates~\cite{Strak91,Bhandari81}, is only about 10\% of the
total dibaryon width. Although the final $\pi NN$ channel has a
different phase-space volume than the $\pi d$ channel, however, in
view of relatively weak influence of the energy dependence of the
total resonance width on final results, we neglect this difference
in the present calculation.

Further, for the partial width $\Gamma_{D_{12} \to \pi d}$ we
employed essentially the same parametrization as for the
$\Delta$-isobar width $\Gamma_{\Delta \to \pi N}$ (up to a factor
$M/W$ which is almost negligible for the dibaryon in the
considered energy region):
\begin{equation}
\label{gpid} \Gamma_{D_{12} \to \pi d}(q) = \Gamma_{D_{12} \to \pi
d} \left(\frac{q}{q_0}\right)^3 \left(\frac{q_0^2 + \Lambda_{\pi
d}^2}{q^2 + \Lambda_{\pi d}^2}\right)^2,
\end{equation}
where $q_0$ is a value of the $\pi d$ relative momentum $q$ at the
energy $\sqrt{s} = M_{D_{12}} = 2.15$ GeV and $\Lambda_{\pi d} =
\tilde{\Lambda}_* = 0.3$ GeV (cf. Eq.~(\ref{gdel})). Given the
fact that the basic hadronic component of the dibaryon
$\mathcal{D}_{12}$ is $N + \Delta$~\cite{Gal14}, one may assume
that the mechanism of the dibaryon decay at the quark level is
essentially the same as that of the $\Delta$ decay, so the above
choice seems to be quite natural. For the vertex $\mathcal{D}_{12}
\to NN$, we used the Gaussian form factor derived on the basis of
the quark shell model~\cite{JPG01K}. In the work~\cite{JPG01K}, a
fit was performed for the $NN$-scattering phase shifts up to the
energies $T_N = 600$ MeV, within the framework of the dibaryon
model for $NN$ interaction. In its simplest form, the model
included pion exchange at large $NN$ distances and intermediate
dibaryon ($\mathcal{D}$) formation at small distances. The
authors~\cite{JPG01K} found the $\mathcal{D} \to NN$ vertex form
factors in various $NN$ partial waves in the form of projections
of the six-quark wave functions onto the $NN$ channel. In the
quark shell model, such projections have the form of the harmonic
oscillator wave functions. So, from the fit of the $NN$ phase
shift in the $^1D_2$ partial wave, the Gaussian (oscillator) form
factor for the $\mathcal{D}_{12} \to NN$ vertex was obtained with
a scale parameter $\alpha = 0.25$ GeV. In the present work, we
took just this value as a first estimate.\footnote{One should bear
in mind that this value may be changed slightly, when one takes as
a background to the dibaryon mechanism not only pion exchange with
intermediate nucleon, but also pion exchange with intermediate
$\Delta$ excitation.} Thus, for the incoming width
$\mathcal{D}_{12} \to NN$, we used the following parametrization:
\begin{equation}
\Gamma_{D_{12} \to NN}(p) \!=\! \Gamma_{D_{12} \to NN}\!
\left(\frac{p}{p_0}\right)^5 \! {\rm
exp}\!\left(-\frac{p^2-p_0^2}{\alpha^2}\right), \label{gdnn}
\end{equation}
where $p_0$ is a value of the $NN$ relative momentum $p$ at
$\sqrt{s} = 2.15$ GeV.

As was mentioned above, the partial width $\Gamma_{D_{12}\to NN}$
is only $\simeq 10 \%$ of the total width $\Gamma_{D_{12}}$, so,
it is reasonable to assume $\Gamma_{D_{12}\to NN} = 10$ MeV. For
the width $\Gamma_{D_{12} \to \pi d}$, there are various estimates
in the literature. The most restrictive estimate $\Gamma_{D_{12}
\to \pi d} / \Gamma_{D_{12}} \lesssim 0.1$ was obtained from the
theoretical analysis of $\pi^+ d$ elastic scattering in the
$\Delta$ region in several independent
works~\cite{Simonov79,Ferreira83}.
%So, one should choose the value
%for $\Gamma_{D_{12} \to \pi d}$ from the above interval. However,
%one should also take into account the higher ratio
%$\Gamma_{D_{12} \to \pi d}/ \Gamma_{D_{12}} \simeq 0.3$ derived
%from PWA for the processes $pp \to pp$, $\pi^+ d \to \pi^+ d$ and
%$\pi^+ d \to pp$~\cite{Strak91}.
Given the high inelasticity of the $\mathcal{D}_{12}$ dibaryon in
the $NN$ channel, it seems natural to assume the $\Gamma_{D_{12}
\to \pi d}$ width to be not less than the $\Gamma_{D_{12} \to NN}$
one. So, for the present calculation, we have taken the value
$\Gamma_{D_{12} \to \pi d} \simeq 0.1 \Gamma_{D_{12}} \simeq
\Gamma_{D_{12} \to NN}  = 10$ MeV.

Now it remains to determine the relative phase $\varphi_{12}$
between the resonance amplitude of the $\mathcal{D}_{12}$
excitation and the ``background'' amplitude given by ONE + $N
\Delta$ mechanisms. Since we found the background processes to
give a strong underestimation of the $pp \to d \pi^+$ cross
section (see Fig.~\ref{fig3}), it is natural first to consider the
case $\varphi_{12} = 0$, corresponding to constructive
interference between the resonance and background contributions.
In fact, it turns out that for the above choice of dibaryon
parameters, the phase $\varphi_{12} \simeq 0$ gives the best
description of the data (from the best fit to PWA data, we have
got $\varphi_{12} = 0.04$).

%It should be noted however that choosing a larger value for
%$\Gamma_{D_{12} \to \pi d}$ (see above) and increasing the phase
%$\varphi_{12}$ simultaneously would not worsen the good agreement
%with empirical data. So, the choice of both these parameters
%should be constrained by other related processes such as elastic
%scattering.

%Thus, if the dibaryon mechanism really makes a significant
%contribution to the reaction $NN \to d\pi$ and its parameters are
%close to those we used in the present study, the resonance
%amplitude should interfere with the background one constructively.

\begin{figure}[!ht]
\begin{center}
\resizebox{0.6\columnwidth}{!}{\includegraphics{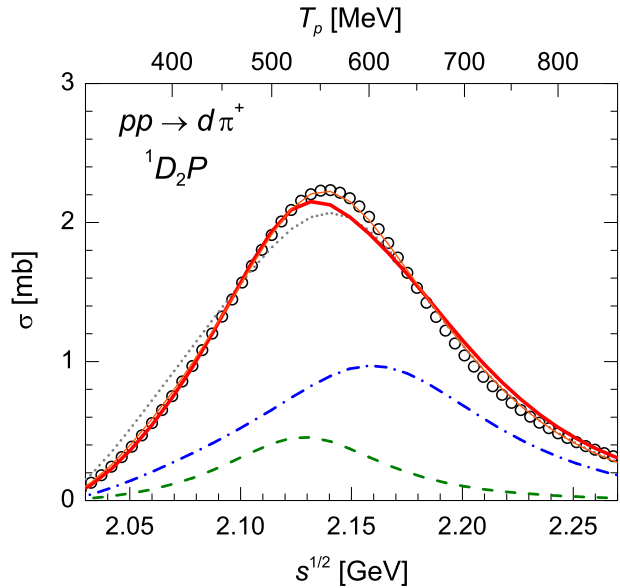}}
\end{center}
\caption{(Color online) Partial cross section of the reaction $pp
\to d\pi^+$ in the $^1D_2P$ channel. The results of calculation
including the conventional ONE + $N\Delta$ mechanisms and an
intermediate dibaryon excitation with the parameter set A (see
Table I), i.e., with no actual fit, is shown by thick solid line.
The individual contributions of the dibaryon excitation (dashed
line) and background (ONE + $N\Delta$) processes (dash-dotted
line) are also shown. The dotted line corresponds to the
calculation with a reduced parameter $\Lambda_{\pi d} = 0.15$~GeV.
The thin solid line shows the results obtained with slightly
modified parameters of the dibaryon mechanism (see Table I, set
B). Open circles correspond to the PWA data (SAID, solution
C500~\cite{SAID,Oh97}).} \label{fig6}
\end{figure}

The results of calculation for the $pp \to d \pi^+$ partial cross
section in the $^1D_2P$ wave with the above fixed dibaryon
parameters (summarized in set A of Table I) and the relative
resonance/background phase $\varphi_{12} = 0$ are shown in
Fig.~\ref{fig6} by thick solid line. We see that the theoretical
curve is in very good agreement with the PWA data at all energies
from threshold up to $\sqrt{s} \simeq 2.3$ GeV. It is important to
emphasize that this result was obtained \emph{without actual fit
of free parameters} for both interfering amplitudes, i.e., the
resonance and background ones.

\begin{table*}[!ht]
\flushleft {\caption{Parameters of dibaryon resonance
$\mathcal{D}_{12}$ (in MeV) used in calculations of the one-pion
production reaction $pp \to d \pi^+$.}}

\begin{tabular*}{1.0\textwidth}{@{\extracolsep{\fill}}ccccccc} \hline
  & $M_{D_{12}}$ & $\Gamma_{D_{12}}$ & $\Gamma_{D_{12} \to pp}$ & $\alpha$
& $\Gamma_{D_{12} \to \pi d}$ & $\Lambda_{\pi d}$
\\ \hline Set A (initial) & 2150 & 110 & 10 & 250
& 10 & 300 \\ Set B (modified) & 2155 & 103 & 10 & 230 & 8.4 & 250
\\ \hline
\end{tabular*}
\end{table*}

At the same time, we found that already a small change in the
basic dibaryon parameters is sufficient to describe the partial
$^1D_2P$ cross section almost perfectly (i.e., in full agreement
with the PWA data), with the same relative phase between the
resonance and background amplitudes $\varphi_{12} = 0$. These
slightly modified parameters are presented in set B of Table I. In
fact, to accurately describe the partial cross section near the
resonance peak, one needs only to increase the dibaryon mass by 5
MeV, while the modification of other parameters improves mainly
the description of the data at higher energies. The result of
calculations with such slightly modified parameters is shown in
Fig.~\ref{fig6} by thin solid line.

Furthermore, we found that the sensitivity of the results to the
dibaryon mass and width is stronger than to the cut-off parameters
in partial widths. However, the deviation of the latter parameters
from the initially adopted values worsens (though not
considerably) the description of the PWA data. For illustration,
we have shown in Fig.~\ref{fig6} also the result of the
calculation with the reduced value $\Lambda_{\pi d} = 0.15$~GeV
(this value corresponds to an assumption of a constant partial
width near the resonance point). Therefore, an accurate
description of the data requires ``fine tuning'' of the model
parameters. Given this fact, it may seem surprising that, by
choosing the parameters from the independent sources and making no
actual fit, we have got a very good agreement with empirical data.
On the other hand, if the parameter values used here are close to
the real ones, then this result becomes quite natural.

\begin{figure}[!ht]
\begin{center}
\resizebox{0.6\columnwidth}{!}{\includegraphics{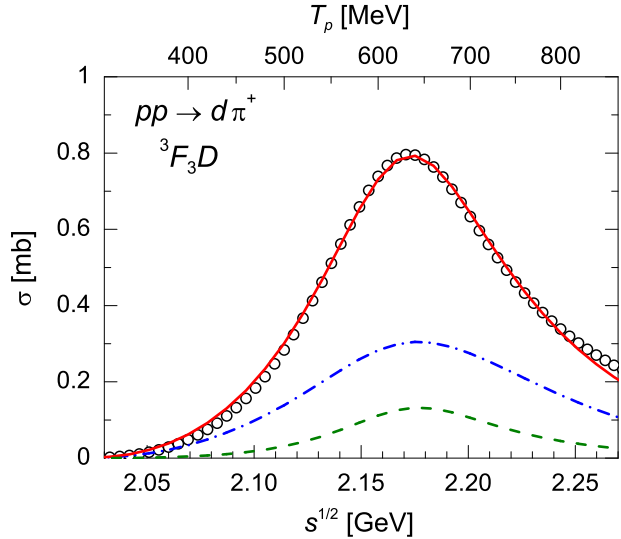}}
\end{center}
\caption{(Color online) Partial cross section of the reaction $pp
\to d\pi^+$ in the $^3F_3D$ channel. The summed contribution of
the conventional ONE + $N\Delta$ mechanisms (dash-dotted line),
the contribution of the dibaryon resonance $\mathcal{D}_{13}^-$
(dashed line) and the full calculation including both mechanisms
(solid line) are shown. Open circles correspond to the PWA data
(SAID, solution C500~\cite{SAID,Oh97}).} \label{fig6-1}
\end{figure}

Let us now consider the $^3F_3D$ partial wave which also gives a
significant contribution to the $pp \to d \pi^+$ cross section.
The results of calculations for the $^3F_3D$ partial cross section
are shown in Fig.~\ref{fig6-1}. Here the conventional ONE +
$N\Delta$ mechanisms with soft cut-off parameters in meson-baryon
vertices (see the previous section) also give about $40$\% of the
partial cross section. An additional contribution can come from a
next (after $\mathcal{D}_{12}$) member of the isovector dibaryon
series, which is usually denoted as $^3F_3(2220)$ (we also will
denote it as $\mathcal{D}_{13}^-$, following the notations of
Ref.~\cite{Dyson64}). This dibaryon has quantum numbers $I(J^P) =
1(3^-)$, mass $M_{D_{13}^-} \simeq 2200$--$2250$ MeV and total
width $\Gamma_{D_{13}^-} \simeq 100$--$200$
MeV~\cite{Makarov82,Yokosawa90}. It was investigated in a number
of works (see, e.g.,~\cite{Hoshizaki78,Bolger81,Akemoto83}, and it
fits well into the classification of isovector dibaryons as a
rotational band for the six-quark system clustered as $[q^4 -
q^2]$~\cite{Mulders80,Kondr87} (see also Sec.~\ref{dibspec}).

We parameterized the amplitude corresponding to the
$\mathcal{D}_{13}^-$ resonance excitation and its total and
partial widths in a similar way as for the $\mathcal{D}_{12}$
resonance (see Eqs.~(\ref{eq-dib})--(\ref{gdnn})), with an
apparent modification due to different angular momenta. When
adding coherently the mechanism of the $\mathcal{D}_{13}^-$
excitation to the summed background (ONE + $N\Delta$)
contribution, one gets a very good description of the $^3F_3D$
partial cross section. However, as the parameters of the
$\mathcal{D}_{13}^-$ dibaryon are known with larger uncertainties
than those of the $\mathcal{D}_{12}$, they cannot be fixed a
priori and still need to be fitted. The results shown in
Fig.~\ref{fig6-1} were obtained with the following set of dibaryon
parameters: $M_{D_{13}^-} = 2200$ MeV, $\Gamma_{D_{13}^-} = 140$
MeV, $\Gamma_{D{13}^- \to pp} \Gamma_{D{13}^- \to \pi d} =
3.5\times 10^{-5}$ GeV$^2$, $\alpha ({}^3F_3) = 0.26$ GeV,
$\Lambda_{\pi d} = 0.3$ GeV. These parameters are consistent with
estimates given in the literature. The relative phase between
resonance and background amplitudes was again fixed to be zero.

\begin{figure}[!ht]
\begin{center}
\resizebox{0.6\columnwidth}{!}{\includegraphics{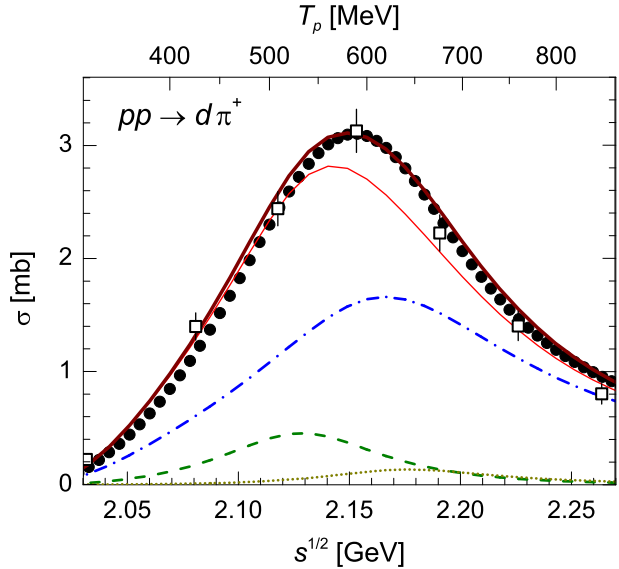}}
\end{center}
\caption{(Color online) Total cross section of the reaction $pp
\to d \pi^+$ with account of the dibaryon resonance in the
$^1D_2P$ partial wave (thin solid line), as well as with two
dibaryon resonances in the $^1D_2P$ and $^3F_3D$ partial waves
included (thick solid line) in comparison with experimental
data~\cite{Shimizu82} (open squares) and the PWA data (SAID,
solution C500~\cite{SAID,Oh97}) (filled circles). The individual
contributions of dibaryons in $^1D_2P$ and $^3F_3D$ partial waves
are shown by dashed and dotted lines, respectively, and the summed
contribution of two conventional processes ONE + $N\Delta$ is
shown by dash-dotted line.} \label{fig7}
\end{figure}

Fig.~\ref{fig7} shows the total $pp \to d \pi^+$ cross section,
calculated with and without intermediate dibaryons excitation
taken into account. Here, in order to estimate the contribution of
other partial waves, except for the dominant one $^1D_2P$, we used
the set of parameters for the dibaryon mechanism, which gives an
accurate reproduction of the partial $^1D_2P$ cross section (set B
in Table~I). It is seen from Fig.~\ref{fig7} that the background
processes yield approximately half of the total cross section (the
result already given in Sec.~\ref{onepic}). When the dibaryon
$\mathcal{D}_{12}$ excitation mechanism is included in the
calculation, the theoretical results are already in a good
agreement with experimental data at low energies. However, at the
energies close to the cross section peak and above there are still
quite visible discrepancies.

Further, when adding both $\mathcal{D}_{12}$ and
$\mathcal{D}_{13}^-$ resonance contributions to the background
(ONE + $N\Delta$) mechanisms, the total $pp \to d \pi^+$ cross
section is described well also at higher energies (see the thick
solid line in Fig.~\ref{fig7}). It is worth noting that including
just these two resonances allowed previously to considerably
improve the description of experimental data also for the
reactions $pp \to pn \pi^+$ and $pp \to pp \pi^0$~\cite{Konig81}.

So, for one-pion production reaction $pp \to d \pi^+$, we have
shown that two lowest (by far dominant) partial waves, where large
contributions from intermediate dibaryons might be expected, are
described by our model very well in a broad energy range. This
result is sufficient for the main objective of the present study,
i.e., to make conclusions about the relative contributions of
dibaryon resonances and background $t$-channel processes. On the
other hand, as is well known, the polarization observables (and
even the differential cross section) are determined by not only
the lowest partial waves, but also by the high partial waves which
give a small ($< 10\%$) contribution to the total cross section.
To describe well these high partial-wave amplitudes, which are
governed basically by the peripheral $t$-channel processes, one
needs a very accurate theoretical model for these processes. Since
even the model based on solving the exact Faddeev-type
equations~\cite{Lamot87} contains ambiguities which affect
strongly the description of observables (e.g., due to account of
small $\pi N$ partial-wave amplitudes, treatment of the off-shell
$\pi N$ amplitude, etc.), here we do not claim the accurate
description of the high partial waves. The detailed calculation of
differential observables and comparison with experimental data is
a subject of our subsequent study.

%\footnote{The best fit of the ${}^3F_3D$ partial cross section was
%obtained with parameters $M_{D}_{13}^- = 2200$ MeV and
%$\Gamma_{D}_{13}^- = 140$ MeV.}

Our final remark here concerns dibaryon resonances in higher $NN$
partial waves $^1G_4$, $^3H_5$, etc., which can only be seen in
particularly sensitive spin-dependent observables (such as the
spin-correlation parameter $C_{LL}$ in elastic $pp$
scattering~\cite{Makarov82}). To study the contributions of such
highly-excited dibaryons to the processes like $pp \to d \pi^+$, a
very accurate description of the background reaction mechanisms is
also required. It seems hardly possible even in a rigorous
multiple-scattering approach, when keeping in mind the vertex form
factors dependence described above.

On the other hand, an appealing opportunity to study these
higher-lying dibaryon resonances is to find such processes where
the relative contributions of background mechanisms at respective
energies are even smaller than in one-pion production. Below, in
Sec.~\ref{twopi}, we consider the \emph{two-pion production}
reactions at intermediate energies from this viewpoint. Before
that, however, it is important to study the dibaryon contributions
(with the same parameters used here to describe one-pion
production) to elastic $NN$ and $\pi d$ scattering.

\subsection{Dibaryon contributions to elastic $NN$ and $\pi d$ scattering}

It is generally known that the overall contribution of short-range
QCD mechanisms to elastic scattering processes is very small,
because elastic scattering occurs mainly in peripheral region and
involves the short-range dynamics only weakly. On the other hand,
when considering large-angle scattering, accompanied by high
momentum transfers, one is likely to see the enhanced
manifestation of quark d.o.f., in particular, the dibaryon
resonances excitation. Indeed, there are numerous indications of
this fact in the literature (see, e.g.,~\cite{Bolger81,Akemoto83},
and also~\cite{PRC10,JPCS12}). However, in the present paper, we
restrict our analysis to the energy dependence of the partial and
total cross sections only.

First of all, let us consider the pure contribution of the
dibaryon $\mathcal{D}_{12}$ excitation to the cross sections of
$pp$ and $\pi^+d$ elastic scattering in partial waves $^1D_2$ and
$^3P_2$, respectively. The $^1D_2$ partial-wave amplitude for $pp$
elastic scattering via the intermediate dibaryon
$\mathcal{D}_{12}$ has the form:
\begin{equation}
A^{(D)}_{pp}(^1D_2) = -\frac{8\pi s}{p}
\frac{\sqrt{2}\Gamma_{D_{12} \to
pp}(s)}{s-M_{D_{12}}^2+i\sqrt{s}\Gamma_{D_{12}}(s)} \label{eq-el}
\end{equation}
and the respective cross section is
\begin{equation}
\sigma^{(D)}_{pp}(^1D_2) = \frac{5}{64\pi
s}\left|A^{(D)}_{pp}(^1D_2)\right|^2.
\end{equation}
Similarly, for the $^3P_2$ partial-wave amplitude and cross
section in $\pi d$ elastic scattering one gets:
\begin{equation}
A^{(D)}_{\pi d}(^3P_2) = -\frac{8\pi s}{q} \frac{\Gamma_{D_{12}
\to \pi d}(s)}{s-M_{D_{12}}^2+i\sqrt{s}\Gamma_{D_{12}}(s)}
\label{eq-el-2}
\end{equation}
and
\begin{equation}
\sigma^{(D)}_{\pi d}(^3P_2) = \frac{5}{48\pi s}\left|A^{(D)}_{\pi
d}(^3P_2)\right|^2.
\end{equation}

\begin{figure}[!ht]
\begin{center}
\resizebox{1.0\columnwidth}{!}{\includegraphics{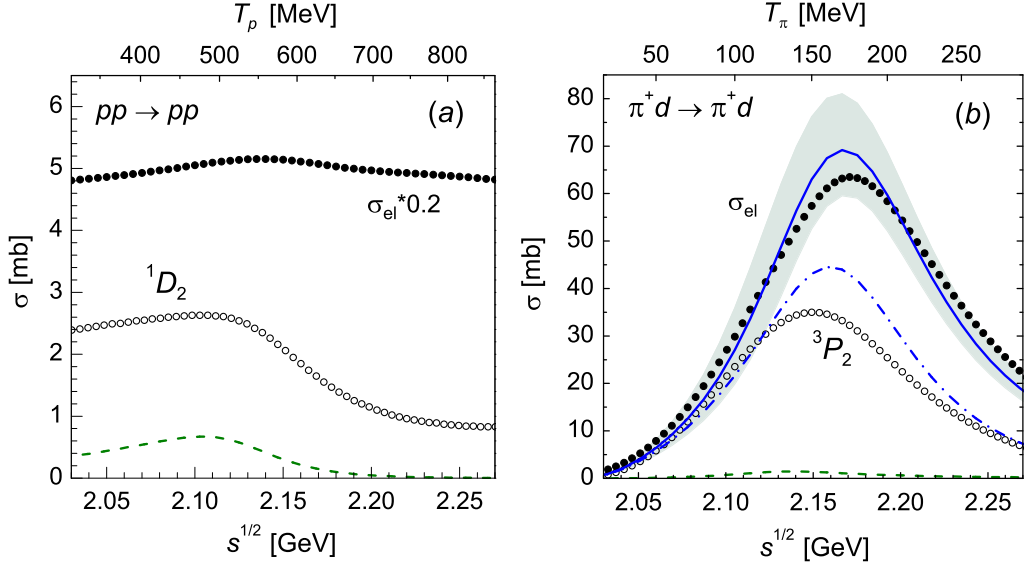}}
\end{center}
\caption{(Color online) Cross sections for $pp$ ($a$) and $\pi^+
d$ ($b$) elastic scattering. The dashed lines correspond to
contributions of the intermediate dibaryon $\mathcal{D}_{12}$
excitation. The open circles show the PWA (SAID~\cite{SAID}) data
for the cross sections in $pp$ $^1D_2$ (solution SP07) and
$\pi^+d$ $^3P_2$ (solution C500) partial waves, and the filled
circles show the respective data for the total elastic cross
sections. The PWA data for the $pp$ total elastic cross section
are multiplied by a factor $0.2$. For $\pi^+d$ scattering, the
dominant contributions of the single $\pi^+N$-scattering mechanism
to the partial $^3P_2$ (dot-dashed line) and the total elastic
(solid line) cross sections are also shown. The shaded area
corresponds to the total elastic $\pi^+d$ cross section resulting
from the coherent superposition of the single scattering and
dibaryon excitation mechanisms with an arbitrary relative phase.}
\label{fig8}
\end{figure}

Fig.~\ref{fig8}~($a$) shows that the dibaryon contribution to the
$^1D_2$ partial cross section of elastic $pp$ scattering is
$\simeq 25\%$ at energies near the resonance peak. However, as the
contribution of the $^1D_2$ partial wave itself is only $10\%$ of
the total elastic $pp$ cross section, the $\mathcal{D}_{12}$
dibaryon contribution to the total elastic cross section will be
$2.5\%$ only. It is interesting to note that the qualitative
behavior of the dibaryon contribution agrees well with the
behavior of the empirical $pp$ cross section in the $^1D_2$
partial wave. This is not unexpected, since we used for the
$\mathcal{D}_{12} \to NN$ vertex the Gaussian form factor obtained
from fitting the $^1D_2$ $NN$-scattering phase shift within the
dibaryon model~\cite{JPG01K}. On the other hand, it is known that
the description of this phase shift within the framework of
conventional meson-exchange models faces a number of problems. In
particular, it requires the introduction of phenomenological
$L^2$-dependent terms in the $NN$ potential (see,
e.g.,~\cite{Wiringa95}). Looking at Fig.~\ref{fig8}~($a$), one may
suppose that, when adding the dibaryon contribution to a
relatively smooth background given by meson-exchange mechanisms
(with soft form factors), one will obtain qualitatively correct
behavior of the $pp$-scattering cross section in the $^1D_2$
partial wave.

Fig.~\ref{fig8}~($b$) shows the partial cross section of elastic
$\pi^+ d$ scattering in the $^3P_2$ wave. An analysis of just this
reaction in its time put the existence of isovector dibaryon
resonances under question, when it became clear that the main
features of experimental data can be explained in terms of the
so-called ``pseudoresonances'', appearing due to an intermediate
$N+\Delta$ excitation~\cite{Simonov79}. The estimate for the
dibaryon partial width $\Gamma_{D_{12} \to \pi d} \lesssim 0.1\,
\Gamma_{D_{12}}$ used in the present work was also obtained from
the analysis of elastic $\pi^+d$
scattering~\cite{Ferreira83,Simonov79}. Our results confirm that
the dibaryon contribution (with the same parameters as were used
to describe one-pion production) to elastic $\pi^+ d$ scattering
even in the $^3P_2$ partial wave is indeed very small ($\simeq
5$\%). Further, since the $^3P_2$ partial wave gives about half
total elastic $\pi^+ d$ cross section near its peak, we obtain the
dibaryon contribution to the total elastic $\pi^+ d$ cross section
to be about $2.5\%$ only, similarly to the case of $pp$ elastic
scattering.

We calculated also the contribution of the standard single
$\pi^+N$-scattering mechanism via an intermediate $\Delta$-isobar
excitation to the partial and total elastic $\pi^+ d$ cross
sections. It is important to stress here that this mechanism,
unlike the similar mechanisms of the intermediate $\Delta$
excitation in the $pp \to d\pi^+$ reaction and in $pp$ elastic
scattering, is very weakly dependent on the cut-off parameter in
the $\pi N \Delta$ vertex, because such vertices here contain the
real pions only, and presence of virtual nucleons produces a very
small effect on the cross sections.

The single-scattering amplitude in the nucleon-spectator
approximation (see Eq.~\ref{spec}) is written as follows:
\begin{equation*}
\mathcal{M}^{(\rm{SS})}_{\lambda_d,\lambda_d'} = -\frac{4}{3}
\,\,\rm{Sp}\int \frac{d^3 P}{(2\pi)^3}
\Psi^*_{d}(\bm{\rho_b},\lambda_d')\sqrt{\frac{\Gamma_{\Delta}(\varkappa)\Gamma_{\Delta}(\varkappa')}{\varkappa^3
\varkappa'^3}}
\end{equation*}
\begin{equation}
\times \frac{16 \pi W_{\Delta}^2(\bm{\varkappa\varkappa'}
 + i \frac{\bm{\sigma}}{2}\bm{\varkappa\times\varkappa'})}{W_{\Delta}^2 - M_{\Delta}^2 + i
W_{\Delta}\Gamma_{\Delta}(W_{\Delta})}\Psi_{d}(\bm{\eta_b},\lambda_d),
\end{equation}
where the momenta are denoted as in Fig.~\ref{fig1}~($b$) (with
the nucleon $1$ and the virtual pion interchanged and the nucleon
$2$ replaced by the incoming deuteron) and the d.w.f. $\Psi_{d}$
is given by Eqs.~(\ref{dwf})--(\ref{e-vect}). There are four
independent helicity amplitudes in $\pi d$ elastic
scattering:\footnote{We use here the same letters to denote the
helicity and partial-wave amplitudes as for the $pp \to d \pi^+$
process since this should not lead to confusion.}
\begin{equation}
\Phi_1 = \mathcal{M}_{1,1}, \quad \Phi_2 = \mathcal{M}_{1,0},
\quad \Phi_3 = \mathcal{M}_{1,-1}, \quad \Phi_4 =
\mathcal{M}_{0,0}.
\end{equation}
Then one has for the total cross section
\begin{equation*}
\sigma(\pi^+ d) = \frac{1}{96\pi
s}\int\limits_{-1}^{1}\Big[2\big(\left|\Phi_1(x)\right|^2 +
\left|\Phi_3(x)\right|^2\big)
\end{equation*}
\begin{equation}
 + 4\left|\Phi_2(x)\right|^2 +
\left|\Phi_4(x)\right|^2\Big] \, dx, \quad x = \rm{cos}(\theta).
\end{equation}
The amplitude in the $^3P_2$ partial wave is expressed through the
helicity amplitudes as
\begin{equation}
A(^3P_2) = \frac{3}{10}\left(\Phi_1^{(2)} + \Phi_3^{(2)}\right) +
\frac{2\sqrt{3}}{5}\Phi_2^{(2)} + \frac{1}{5}\Phi_4^{(2)},
\end{equation}
where
\begin{equation}
\Phi_i^{(J)} =
\int\limits_{-1}^{1}d^{(J)}_{\lambda_d,\lambda_d'}(x)\Phi_i(x) \,
dx.
\end{equation}

Our calculations have shown, in agreement with results of many
previous works~\cite{Simonov79,Ferreira83,VdV77}, that the
single-scattering mechanism gives the by far dominating
contribution to the $\pi^+ d$ elastic cross sections, both partial
and total. We found also that possible interference between the
dibaryon and single-scattering contributions can give a scatter
$\pm 12\%$ (depending on a relative phase) in the total elastic
$\pi^+ d$ cross section, while the cross section shape, after
adding the dibaryon contribution, remains practically unchanged
(see Fig.~\ref{fig8}~($b$)). One should further take into account
that the contribution of multiple scattering processes to the
elastic $\pi^+d$ cross section is $\simeq 20\%$~\cite{Ferreira77},
which is significantly higher than that of the dibaryon mechanism.
As a result, the effect of intermediate dibaryon excitation
appears to be hardly visible in the total elastic cross sections
of $pp$ and $\pi^+ d$ scattering.

Thus, one can conclude from the above analysis that the model
which includes the background (or pseudoresonance) meson-exchange
processes and the intermediate dibaryon excitation mechanisms with
reasonable parameters allows a good description of the one-pion
production reaction $NN \to d \pi$ in a wide energy range and at
the same time does not contradict the empirical data for elastic
$NN$ and $\pi d$ scattering. One should however bear in mind the
strong parameter dependence of such a model description,
especially in the background $t$-channel mechanisms. In the next
section we consider more clear manifestations of intermediate
dibaryon resonances in the two-pion production processes, where
the conventional meson-exchange contributions are expected to be
smaller than those in one-pion production, due to the higher
momentum transfers.

\section{Dibaryon resonances in two-pion production}
\label{twopi}

\subsection{Reaction $pn \to d (\pi\pi)_0$: isoscalar/isovector transition}
In reactions of two-pion production, where the initial $NN$ pair
merges into the final deuteron, there are two possible assignments
for the total isospin $I = 1$ and $0$. The most interesting case
is a purely isoscalar process $pn \to d(\pi\pi)_{I=0}$, where the
famous ABC effect~\cite{Abashian60,Booth61}, i.e., a strong
enhancement in the yield of pion pairs near the $2\pi$ threshold,
was observed~\cite{Bash09} (see also an older inclusive
experiment~\cite{Plouin78}). In recent works of the CELSIUS/WASA
and then WASA@COSY Collaborations~\cite{Bash09,Adl11} the ABC
effect was associated with generation of a dibaryon resonance
$\mathcal{D}_{03}(2380)$ with $I(J^P) = 0(3^+)$, originally
predicted by Dyson and Xuong~\cite{Dyson64} and then studied in
many theoretical and experimental works (see,
e.g.,~\cite{Kamae77,Kamae77-2,Goldman89,Valcarce01,Mota02,Gal13}).
In fact, it is the only isoscalar dibaryon resonance (except for
the deuteron~\cite{Dyson64}) reliably established for today. The
authors~\cite{Adl11} found that the total cross section of the
$2\pi$-production reaction $pn \to d \pi^0\pi^0$ in the energy
range $T_p = 1$--$1.2$ GeV is predominantly determined by
excitation of the intermediate $\mathcal{D}_{03}$ resonance, while
the contribution of background processes (mainly excitation of an
intermediate $\Delta$--$\Delta$ state via a $t$-channel meson
exchange~\cite{Risser73}) is relatively small and does not exceed
$10$\% near the cross section maximum (at $\sqrt{s} = 2.38$~GeV or
$T_p = 1.14$~GeV).
%\footnote{It is important to note that this result was
%obtained using isospin relations and the previous
%calculations~\cite{Kren} for the isovector reaction $pp \to d
%\pi^+\pi^0$, normalized to experimental data. So, if the
%$t$-channel $\Delta\Delta$ process does not entirely determine the
%cross section of the isovector reaction in this energy region (see
%the next subsection), then the background contribution to the
%isoscalar process $pn \to d(\pi\pi)_{0}$ will be even smaller.}
Therefore, the calculations of the $2 \pi$-production reactions in
the isoscalar $NN$ channel based on $\mathcal{D}_{03}$-resonance
excitation only (i.e., without inclusion of the $t$-channel
background processes) can be regarded as a good approximation, at
least near the resonance peak.

In our previous work~\cite{PRC13}, we considered two decay modes
for the resonance $\mathcal{D}_{03}$(2380) into the $d \pi \pi$
channel, which follow directly from the dibaryon model for $NN$
interaction~\cite{AP10K}: (i) through emission of a light scalar
$\sigma$ meson and (ii) via an intermediate state
$\mathcal{D}_{12}(2150) + \pi$. In other words, we assumed that
while the isovector dibaryons cannot be excited directly in the
isoscalar $NN$ collisions, these dibaryons may be produced in the
intermediate subsystem $\pi NN$, i.e., after the one-pion
emission. Now we can verify this assumption independently,
comparing the parameters of the isovector dibaryon
$\mathcal{D}_{12}$ in one- and two-pion production processes. We
actually used in the present paper the same values for the
$\mathcal{D}_{12}$ mass and width, as in the previous calculations
of $2 \pi$ production~\cite{PRC13}. However, the cut-off parameter
$\Lambda_{\pi d}$ in the partial decay width $\mathcal{D}_{12} \to
\pi d$ was chosen here to be $0.3$ GeV, while in~\cite{PRC13} the
smaller value $0.15$ GeV was used. As was shown above, the last
value of $\Lambda_{\pi d}$ leads to a slight disagreement with the
empirical data for the one-pion production at low energies (see
Fig.~\ref{fig6}), though due to uncertainties in our calculation
for background processes, this discrepancy can hardly be
considered to be significant. Below we will test the sensitivity
of the $2 \pi$-production cross sections to the value of
$\Lambda_{\pi d}$.

In the dibaryon model for $2\pi$ production~\cite{PRC13}, the
amplitude for the reaction $pn \to d\pi^0 \pi^0$ can be written as
follows:
\begin{equation}
\label{M} \mathcal{M}_{\lambda_p,\lambda_n,\lambda_d} =
\frac{\sum\limits_{\lambda_3}
\mathcal{M}^{(D_{03})}_{\lambda_p,\lambda_n,\lambda_3}\left[\mathcal{M}^{(\sigma)}_{\lambda_3,\lambda_d}
+
\mathcal{M}^{(D_{12})}_{\lambda_3,\lambda_d}\right]}{s-M_{D_{03}}^2+i\sqrt{s}\Gamma_{D_{03}}(s)}.
\end{equation}
When choosing the $z$ axis to be parallel to the initial c.m.
momentum ${\bf p}$, the dibaryon $\mathcal{D}_{03}$ formation
amplitude takes the form
\begin{equation}
\mathcal{M}^{(D_{03})}_{\lambda_p,\lambda_n,\lambda_3} =
\sqrt{5}p^2 F_{pn \to D_{03}}
C^{3\lambda_3}_{1\lambda_320}C^{1\lambda_3}_{\frac{1}{2}\lambda_p\frac{1}{2}\lambda_n},
\label{eq42}
\end{equation}
$C^{J\Lambda}_{s_1\lambda_1 s_2\lambda_2}$ being the
Clebsch--Gordan coefficients. In its turn, for the dibaryon decay
amplitudes, one gets the following expressions:
\begin{equation}
\mathcal{M}^{(\sigma)}_{\lambda_3,\lambda_d} = \frac{F_{D_{03} \to
d \sigma}F_{\sigma \to
\pi\pi}}{M_{\pi\pi}^2-m_{\sigma}^2+iM_{\pi\pi}\Gamma_{D_{03}}(M_{\pi\pi}^2)}
C^{3\lambda_3}_{1\lambda_d 2\mu}\mathcal{Y}_{2\mu}({\bf p}_d,{\bf
p}_d),
\end{equation}
\begin{equation*}
\mathcal{M}^{(D_{12})}_{\lambda_3,\lambda_d} =
\sqrt{\frac{6}{5}}\frac{F_{D_{03} \to D_{12} \pi_1}F_{D_{12} \to d
\pi_2}}{M_{d\pi_2}^2-M_{D_{12}}^2+iM_{d\pi_2}\Gamma_{D_{12}}(M_{d\pi_2}^2)}
\end{equation*}
\begin{equation}
\times C^{3\lambda_3}_{1\lambda_d 2\mu}\mathcal{Y}_{2\mu}({\bf
p}_{\pi_1},{\bf p}_{d\pi_2}) + (\pi_1 \leftrightarrow \pi_2),
\label{eq44}
\end{equation}
where $\mathcal{Y}_{2\mu}({\bf p}_{1},{\bf p}_{2})$ denote the
solid spherical harmonics, expressed as functions of two momentum
vectors, and $\mu = \lambda_3 - \lambda_d$. The derivation of
Eqs.~(\ref{eq42})--(\ref{eq44}) can be done straightforwardly
using the canonical or the non-relativistic spin tensor
formalism~\cite{Chung14}. The detailed derivation can be found
in~\cite{MyPhD}.

The vertex functions are related to the partial decay widths as
follows:
\begin{equation}
F_{R \to ab}(p_{ab}) = M_{ab}\sqrt{\frac{8\pi\Gamma_{R \to
ab}^{(l)}(p_{ab})}{(p_{ab})^{2l+1}}}.
\end{equation}
Further, for the partial decay widths with meson emission, we
chose the standard parametrization
\begin{equation}
\label{G} \Gamma^{(l)}_{R \to ab}(p) = \Gamma^{(l)}_{R \to ab}
\left(\frac{p}{p_0}\right)^{2l+1} \left(\frac{p_0^2 +
\Lambda_{ab}^2}{p^2+\Lambda_{ab}^2}\right)^{l+1},
\end{equation}
while for the $pn \to \mathcal{D}_{03}$ vertex, we used the
Gaussian form factor, according to the dibaryon model for $NN$
interaction~\cite{JPG01K,IJMP02K}. In this case, the
$\mathcal{D}_{03}$ decay width into $np$ channel has the form
similar to Eq.~(\ref{gdnn}). Parameters $\Lambda_{ab}$ were fixed
by a condition of a constant width near the resonance point, so we
found $\Lambda_{\sigma d} = 0.18$, $\Lambda_{\pi\pi} = 0.09$,
$\Lambda_{\pi D_{12}} = 0.12$ and $\Lambda_{\pi d} = 0.15$ GeV.
For the latter parameter, we used also a higher value
$\Lambda_{\pi d} = 0.3$ GeV which describes better the one-pion
production process $pp \to d \pi^+$ (see Sec.~\ref{onepid}).

The differential distribution on the invariant mass of two
particles $b$ and $c$ can be found from the formula
\begin{equation}
\label{dsbc} \frac{d\sigma}{dM_{bc}} =  \frac{1}{(4\pi)^{5} p s}\!
\int{\!\! \int{\! p_a p_{bc} d\Omega_a d\Omega_{bc} \,
\overline{|\mathcal{M}({\bf p}_a,{\bf p}_{bc})|^2}}},
\end{equation}
where ${\bf p}_a$ is a c.m. 3-momentum of the particle $a$, ${\bf
p}_{bc}$ the particle $b$ momentum in c.m.s. of two particles $b$
and $c$, and the line over the matrix element squared stands for
averaging over the initial and summing over the final spin states.
Then one gets for the total cross section:
\begin{equation}
\label{st} \sigma = \int\limits_{m_b+m_c}^{\sqrt{s}-m_a} \!\! {d
M_{bc} \frac{d\sigma}{dM_{bc}}}.
\end{equation}

\begin{figure}[!ht]
\begin{center}
\resizebox{0.6\columnwidth}{!}{\includegraphics{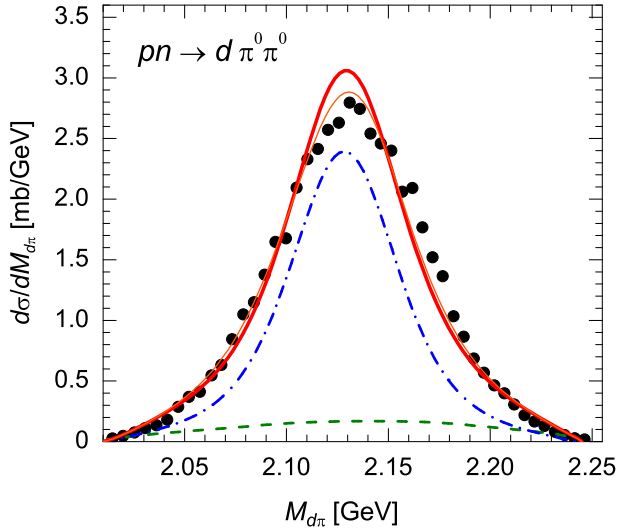}}
\end{center}
\caption{(Color online) Distribution on the invariant mass of the
$d \pi$ system in the reaction $pn \to d\pi^0\pi^0$ at $\sqrt{s} =
2.38$ GeV. Thick solid line corresponds to the parameters of the
isovector dibaryon $\mathcal{D}_{12}$ fixed in Sec.~\ref{onepid}
(see Table~1, set A) and thin solid line was obtained using
slightly modified parameters (set B). The individual contributions
of the two decay modes of the $\mathcal{D}_{03}$ dibaryon, i.e.,
via an intermediate state $\mathcal{D}_{12} + \pi$ (dot-dashed
line) and through a light scalar $\sigma$-meson emission (dashed
line), calculated with the first set of parameters, are also
shown. The filled circles correspond to the WASA@COSY experimental
data~\cite{Adl11} renormalized according to~\cite{Adl13-iso}.}
\label{fig9}
\end{figure}

The significance of the $\sigma$-production mechanism for
description of the $M_{\pi\pi}$ spectrum and the ABC effect was
shown in Ref.~\cite{PRC13}. Here we concentrate on the second
mechanism, i.e., $\mathcal{D}_{03} \to \mathcal{D}_{12} + \pi \to
d + \pi\pi$, which contains the transition between isoscalar and
isovector dibaryons. The most sensitive quantity to the parameters
of the $\mathcal{D}_{12}$ dibaryon is the distribution on the
invariant mass $M_{d \pi}$, while the total cross section as well
as the $M_{\pi\pi}$ distribution are only slightly renormalized
when changing the $\mathcal{D}_{12}$ parameters. Fig.~\ref{fig9}
shows the $M_{d \pi}$ distribution in the reaction $pn \to d \pi^0
\pi^0$ at the peak energy $\sqrt{s} = 2.38$ GeV. Contrary to the
supposition made at the beginning of this section, we found this
distribution to be quite weakly dependent on the parameter
$\Lambda_{\pi d}$. An increase of $\Lambda_{\pi d}$ from $0.15$ to
$0.3$ GeV leads to only a small narrowing and increasing the peak
by about $10\%$, thus worsening somewhat the agreement with
experiment. However, when using a slightly modified set of
parameters for the dibaryon $\mathcal{D}_{12}$, i.e., $M'_{D_{12}}
= 2155$ MeV, $\Gamma'_{D_{12}} = 103$ MeV and $\Lambda'_{\pi d} =
0.25$ GeV (see set B in Table I), which gives an accurate
description of the partial $^1D_2P$ cross section in the one-pion
production process, we got an almost accurate description of the
$M_{d \pi}$ distribution in $2 \pi$ production as well (cf. thin
solid lines in Figs.~\ref{fig6} and \ref{fig9}), the main
improvement being given again by a small increase in the
$\mathcal{D}_{12}$ mass. This indicates the possibility of a very
good \emph{simultaneous description of the independent empirical
data} for one- and two-pion production processes with the same
realistic parameters of the $\mathcal{D}_{12}$ dibaryon.

On the other hand, we found that changing the $\mathcal{D}_{12}$
mass by $10$--$20$ MeV does not lead to any shift of the resonance
peak position in the $M_{d \pi}$ spectrum in the $2 \pi$
production process, in contrast to the cross section of the
one-pion production reaction $NN \to d \pi$. This reflects the
fact that all final distributions in the reaction $pn \to d
(\pi\pi)_0$ must be symmetrized over two outgoing pions. As a
consequence of this symmetry, simultaneous changes in two
individual distributions for each pion largely cancel each other
and are therefore weakly reflected in a final (observed)
distribution. Given this fact, it is not surprising that the
$M_{d\pi}$ distribution is well reproduced also by a mechanism
$\mathcal{D}_{03} \to \Delta\Delta$~\cite{Bash09,Adl11}, without
the formal account of the $\mathcal{D}_{12}$ dibaryon. In this
case, each individual-pion distribution on the $d \pi$ invariant
mass peaks near the $N\Delta$ threshold located 20 MeV above the
$\mathcal{D}_{12}$ mass. However, the final symmetrized $M_{d
\pi}$ distribution turns out to be almost the same as in case of
intermediate $\mathcal{D}_{12}$ excitation. This makes difficult
to disentangle the two $\mathcal{D}_{03}$ decay routes, i.e.,
$\mathcal{D}_{03} \to \mathcal{D}_{12} + \pi$ and
$\mathcal{D}_{03} \to \Delta\Delta$. From a general viewpoint, of
course, one has to take both these routes into account. However,
it is known from numerous six-quark microscopic
calculations~\cite{Garc97,Yuan99,Wong99} that the wave function of
the $\Delta$--$\Delta$ system in the $I(J^P) = 0 (3^+)$ channel
(corresponding to the $\mathcal{D}_{03}$ resonance) has a very
small mean square radius $r_{\Delta\Delta} \simeq 0.7$--$0.9$ fm,
that is, two $\Delta$ isobars in this state are almost completely
overlapped with each other. Therefore it seems natural to assume
that the main hadronic component of the $\mathcal{D}_{03}$
dibaryon, i.e., $\Delta\Delta$, is not a physical system of two
isolated $\Delta$ isobars, which is highly unstable
($\Gamma_{\Delta\Delta} \simeq 2 \Gamma_{\Delta} \simeq 235$ MeV),
but only a specific $6q$-configuration with quantum numbers of the
$\Delta\Delta$ system. It is also confirmed by an experimental
observation that the $\mathcal{D}_{03}$ resonance width is much
smaller than the total width of two isolated $\Delta$ isobars:
$\Gamma_{D_{03}} \simeq 70$ MeV $\ll \Gamma_{\Delta\Delta}$. In
this context, the independent pion decay of two strongly
overlapped $\Delta$ isobars assumed in~\cite{Bash09,Adl11} seems
not fully justified from physical point of view. Therefore, it
seems more reasonable to suggest that at least one of the final
pions should be emitted from a dibaryon state, i.e., from the
compact $6q$ object surrounded by meson fields, but not from an
isolated $\Delta$ isobar. One also needs to take into account that
the width of the intermediate $\mathcal{D}_{12} + \pi$ state is
about half the width of the $\Delta + \Delta$ state, and thus the
lifetime of the first is two times longer. So, the decay of the
$\mathcal{D}_{03}$ resonance via the intermediate state
$\mathcal{D}_{12} + \pi$ rather than $\Delta + \Delta$ is likely
to be regarded as the dominant one, although both above dibaryons
can formally be described in terms of intermediate $\Delta\Delta$
and $N\Delta$ states~\cite{Gal13}.

\begin{figure}[!ht]
\begin{center}
\resizebox{0.6\columnwidth}{!}{\includegraphics{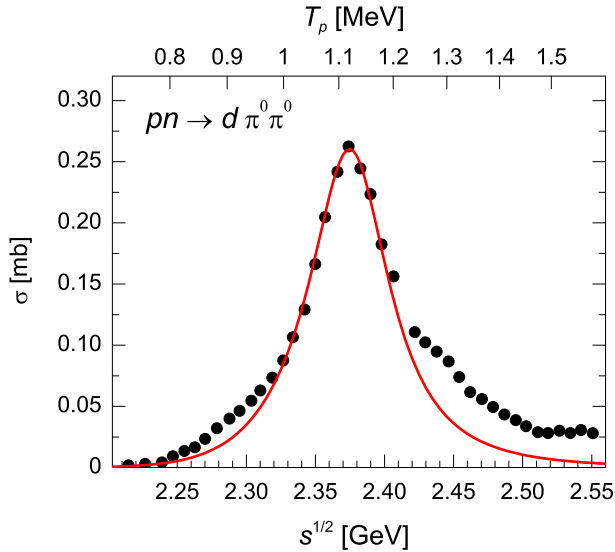}}
\end{center}
\caption{(Color online) Total cross section of the reaction $pn
\to d\pi^0\pi^0$. The solid line shows the calculation in the
dibaryon model which includes excitation of the isoscalar
resonance $\mathcal{D}_{03}(2380)$ with the total width
$\Gamma_{D_{03}} = 70$ MeV. Filled circles correspond to the
WASA@COSY experimental data~\cite{Adl11} renormalized according
to~\cite{Adl13-iso}.} \label{fig10}
\end{figure}

The total cross section of the reaction $pn \to d\pi^0\pi^0$,
going through the formation of the intermediate dibaryon
$\mathcal{D}_{03}$(2380) with a total width $\Gamma_{D_{03}} = 70$
MeV, is shown in Fig.~\ref{fig10}. We obtained a very good
agreement with experimental data at energies close to the
resonance peak using the Gaussian form factor in the $pn \to
\mathcal{D}_{03}$ vertex with a scale parameter
$\alpha(\mathcal{D}_{03}) = 0.35$ GeV which turned out to be
larger than that for the $pp \to \mathcal{D}_{12}$ vertex
$\alpha(\mathcal{D}_{12}) = 0.25$ GeV. This result seems quite
natural because the isoscalar resonance $\mathcal{D}_{03}$,
according to quark-model estimates (see, e.g.,~\cite{Garc97}), is
characterized by a smaller radius than the isovector resonance
$\mathcal{D}_{12}$.

On the other hand, Fig.~\ref{fig10} shows rather large
discrepancies between our theoretical calculation and experimental
data beyond the resonance peak. Particularly strong deviations are
observed at energies $\sqrt{s} \gtrsim 2.43$ GeV. It is well known
however~\cite{Adl11} that a significant contribution at these
energies can be given by the conventional mechanism based on
$t$-channel excitation of the intermediate $\Delta$--$\Delta$
system~\cite{Risser73,BarNir75}, being produced near its threshold
$(\sqrt{s})_{\Delta\Delta} = 2.46$ GeV. An additional enhancement
of the cross section near the $\Delta\Delta$ threshold can come
from interference between the background $t$-channel process and
the resonance $\mathcal{D}_{03}$ contribution which is though
relatively small but still non-zero in this energy region (see
Fig.~\ref{fig10}).

%It is important to add here that the new measurements for the
%total cross section of the reaction $pn \to pn \pi^0\pi^0$ at the
%energies $T_p \sim 1$~GeV has been published recently by the
%WASA@COSY Collaboration~\cite{Adl14-pn}. The experiment clearly
%shows a significant strengthening of the cross section at $T_p
%\simeq 1.1$ GeV corresponding to excitation of the isoscalar
%resonance $\mathcal{D}_{03}(2380)$. Besides that, the dibaryon
%$\mathcal{D}_{03}$ excitation has been confirmed recently in the
%PWA of elastic $np$ scattering~\cite{Adl14-el}, and the dibaryon
%parameters found there proved to be in good agreement with those
%derived from experimental data on $2\pi$ production.

The next important and nontrivial step towards establishing a
connection between different hadronic processes and intermediate
dibaryon resonances, is searching for isovector dibaryon signals
in the processes of two-pion production in $pp$ collisions.

\subsection{Isovector dibaryon signals in $2\pi$ production in $pp$ collisions}

To the present authors' knowledge, excitation of intermediate
dibaryon resonances in two-pion production processes in
\emph{isovector} $NN$ channels, like $pp \to d \pi^+\pi^0$, $pp
\to pp \pi^0\pi^0$, etc., has not yet been considered in the
literature. In fact, the mass of the basic isovector dibaryon
$\mathcal{D}_{12}$ lies just at the $2\pi$-production threshold
$(\sqrt{s})_{\rm NN \pi\pi} = (2m_p + 2m_{\pi^0}) \approx 2.15$
GeV, so its decay with two-pion emission is very unlikely. However
the higher-lying isovector dibaryons found in $pp$ elastic
scattering in partial waves $^3F_3$, $^1G_4$,
etc.~\cite{Makarov82,Yokosawa90}, if they really exist, should
decay into $d \pi\pi$ and $NN\pi\pi$ channels with a higher
probability. Thus, the dibaryon $\mathcal{D}_{13}^-(2220)$
($^3F_3$) should be excited in $pp$ collisions at energies $T_p =
(M_D^2/2m_p - 2m_p) \approx 750$ MeV, the dibaryon
$\mathcal{D}_{14}(2430)$ ($^1G_4$)
--- at $T_p \approx 1.3$ GeV, etc. The possibility of finding the signals
of these dibaryons in the $2\pi$-production cross sections is
determined mainly by the relative contributions of the resonance
and background processes. However, as was already outlined above,
one might expect the contributions of the background
meson-exchange mechanisms (with relatively soft vertex cut-offs)
to the two-pion production to be significantly less than to
elastic scattering or one-pion production, since $2\pi$ production
is generally accompanied by larger momentum transfers.

The conventional mechanisms of $2\pi$ production in $pp$
collisions at energies $T_p \sim 1$ GeV are based on $t$-channel
excitation of the intermediate ``pseudoresonance'' systems $NR$
($T_p \simeq 0.9$--$1.1$ GeV), where $R$ is the Roper resonance
$N^*(1440)$, and $\Delta\Delta$ ($T_p \simeq 1.3$--$1.4$ GeV). So,
one may assume that the above meson-exchange mechanisms can
interfere with the true resonance ones based on formation of
intermediate isovector dibaryons. It should be borne in mind that
the mass of the $\mathcal{D}_{14}(2430)$ dibaryon lies very close
to the $\Delta\Delta$-excitation threshold
$(\sqrt{s})_{\Delta\Delta} = 2.46$ GeV, so that the contribution
of this dibaryon resonance may be difficult to separate from the
contribution of the conventional $t$-channel $\Delta\Delta$
process. One faces here in principle the same problems as in
determining the relative contributions of the true resonance
$\mathcal{D}_{12}(2150)$ and the ``pseudoresonance'' $N\Delta$
when studying the one-pion production processes. However, it
should be emphasized once again that, due to shorter-range nature
of the $2\pi$-production processes, formation of a compact
six-quark object (dibaryon) in this case can have a higher
probability than the much more peripheral $t$-channel meson
exchange, if the latter is calculated with the use of realistic
(soft) vertex form factors. On the other hand, the mass of the
resonance $\mathcal{D}_{13}^-(2220)$ lies 100--200 MeV below the
excitation energies of the $NR$ system, so that the signal of this
dibaryon at energies $T_p \simeq 700$--$800$ MeV can in principle
be seen above the background, although the total cross sections of
$2\pi$ production are very small in this energy region (2--3
$\mu$b only).

\begin{figure}[!ht]
\begin{center}
\resizebox{0.6\columnwidth}{!}{\includegraphics{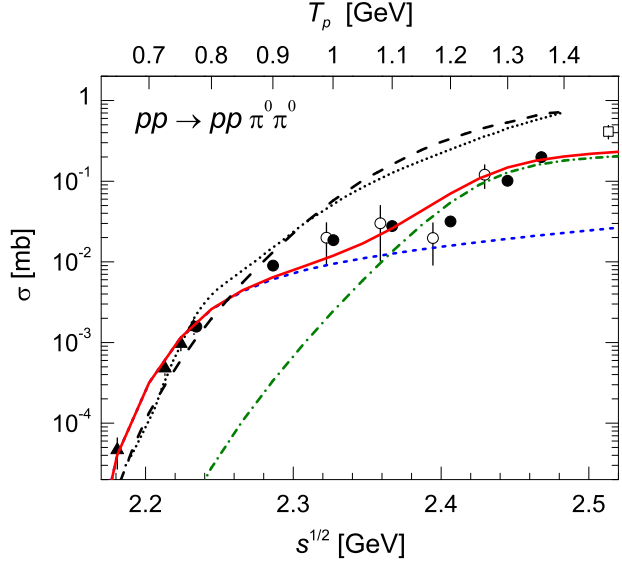}}
\end{center}
\caption{(Color online) Total cross section of the reaction $pp
\to pp \pi^0 \pi^0$. The solid line shows the calculation in a
model including the excitation of two intermediate dibaryon
resonances $\mathcal{D}_{13}^-(2220)$ and
$\mathcal{D}_{14}(2430)$. The individual contributions of the two
resonance mechanisms are shown by short-dashed and dash-dotted
lines, respectively. The dotted line corresponds to the Valencia
model calculations~\cite{Alvarez98} with account of $t$-channel
excitation of intermediate $NR$ and $\Delta\Delta$ states. The
long-dashed line shows the calculations of Cao et al.~\cite{Cao10}
which include all nucleon resonances with masses up to $1.72$ GeV.
The CELSIUS/WASA experimental data are shown by filled symbols and
the older bubble-chamber data
--- by open symbols (see~\cite{Skorodko11} and references
therein). } \label{fig11a}
\end{figure}

Let's consider the reaction $pp \to pp \pi^0 \pi^0$, for which
high-statistics experimental data in a broad energy range
exist~\cite{Skorodko11}. The most intriguing feature of the total
cross section of just this reaction (contrary to the other $NN \to
NN\pi\pi$ channels) is a ``shoulder'' at energies $T_p = 1$--$1.2$
GeV (see Fig.~\ref{fig11a}), which is followed by a rapid increase
as the energy approaches the $\Delta\Delta$ excitation threshold.
The conventional model of the Valencia group~\cite{Alvarez98},
based on $t$-channel excitation of the intermediate states $NR$
and $\Delta\Delta$, even with high cut-off parameters in
meson-baryon vertices taken from the Bonn $NN$-potential
model~\cite{Machl87}, \footnote{Note that $t$-channel
$\Delta\Delta$ process is even more sensitive to the cut-off
parameter in the $\pi N \Delta$ vertex than the $N\Delta$
mechanism in one-pion production since the first process contains
\emph{two} such vertices.} does not reproduce the observed
behavior of experimental data because the theoretical cross
section in this model increases uniformly with rising energy.
Aside from the Valencia model, the most recent attempt to describe
two-pion production solely by $t$-channel meson-exchange
mechanisms was made by Cao et al.~\cite{Cao10}. Their model
includes all nucleon resonances with masses up to $1.72$ GeV and
gives a good reproduction of the total and also some measured
differential cross sections for six $NN \to NN \pi\pi$ channels up
to beam energies of $2.2$ GeV, however, also fails in reproduction
of the data for the $pp \to pp \pi^0 \pi^0$ at energies $T_p
\simeq 1.1$--$1.2$ GeV. In fact, the reaction $pp \to pp \pi^0
\pi^0$ is the only two-pion production process in $pp$ collisions
where an almost purely isoscalar $\pi\pi$ pair (with a small
isotensor admixture) is produced. In this channel, the isovector
dibaryon contributions should be enhanced due to intermediate
$\sigma$-meson production (though this enhancement would be less
significant than for the isoscalar dibaryon $\mathcal{D}_{03}$ ---
see Sec.~\ref{pppn}). Other $\pi\pi$-production channels in $pp$ collisions contain
large contributions from isovector $\pi\pi$ pairs which can be
described basically by the background processes ($t$-channel
$\Delta\Delta$ excitation, etc.)~\cite{Cao10}. On the other hand,
the experimental data for $pp \to pp \pi^0 \pi^0$ reaction can be
qualitatively explained by assuming the dominant contribution of
the two known dibaryon candidates: $\mathcal{D}_{13}^-(2220)$ at
$T_p \simeq 750$ MeV and $\mathcal{D}_{14}(2430)$ at $T_p \simeq
1.3$ GeV. In this case, the total cross section of the reaction
$pp \to pp (\pi\pi)_0$ can be described by the formula
\begin{equation}
\sigma = \sum\limits_{J=3,4}\frac{\pi(2J+1)}{p^2}\frac{2s
\Gamma_{D_J}^{(i)}(s)\Gamma_{D_J}^{(f)}(s)}{(s-M_{D_J}^2)^2+s\Gamma_{D_J}^2(s)},
\label{eq-2r}
\end{equation}
where $\Gamma_{D_J}^{(i)}(s)$ and $\Gamma_{D_J}^{(f)}(s)$ denote
the partial widths of the resonance with a total angular momentum
$J$ for the incoming ($pp$) and outgoing ($pp\pi^0\pi^0$)
channels. As a first approximation, the total widths of the two
resonances can be assumed constant and equal to $\Gamma_{D_J} =
150$ MeV. The incoming partial widths $\Gamma_{D_J \to pp}$ can
also be considered constant in this energy range, but for
comparison of the theoretical calculations with experimental data
near the $2\pi$-production threshold one needs to take into
account somehow the energy dependence of the outgoing widths
$\Gamma_{D_J \to pp\pi^0\pi^0}$. In fact, they are proportional to
the factor $(s-4(m + m_{\pi})^2)^n$, where the exponent $n$, in
general, depends on the reaction dynamics. We found that the
energy dependence of the total cross section for the reaction $pp
\to pp \pi^0 \pi^0$ in the near-threshold region can be reproduced
well with $n = 4$. The above factor also significantly distorts
the Breit--Wigner form of the energy distributions corresponding
to the suggested resonances.

The results of calculations using Eq.~(\ref{eq-2r}), as well as
the predictions of the Valencia~\cite{Alvarez98} and Beijing~\cite{Cao10} models are shown in Fig.~\ref{fig11a}.
In order to reproduce the experimental cross section in the
vicinity of the incoming proton energies $T_p = 750$ MeV and $1.3$
GeV, corresponding to the maximal excitation of the above two
isovector resonances, their widths should satisfy the relations:
$\Gamma^{(i)}\Gamma^{(f)}/\Gamma^2 \simeq 2.2 \times 10^{-5}$ and
$2.7 \times 10^{-3}$ for the $\mathcal{D}_{13}^-$ and
$\mathcal{D}_{14}$ dibaryons, respectively. If to suggest the
incoming (elastic) partial widths of these dibaryons to be $\simeq
10\%$ of their total widths (as for the $\mathcal{D}_{12}$
resonance) then we obtain the following estimates for the
branching ratios of the $pp \pi^0\pi^0$ channel: $\Gamma_{D_
{13}^-\to pp \pi^0 \pi^0} / \Gamma_{D_{13}^-} \simeq 0.02\%$ and
$\Gamma_{D_{14} \to pp \pi^0 \pi^0} / \Gamma_{D_{14}} \simeq 3\%$.
These estimates, as follows from Eq.~(\ref{eq-2r}), do not depend
on neither the absolute values of the partial and total widths,
nor the parametrization of their energy dependence. So, one can
see that a very small fraction of the $pp \pi^0 \pi^0$ channel in
the isovector dibaryon decay widths is sufficient to describe the
total cross section of the reaction $pp \to pp \pi^0 \pi^0$ in
terms of intermediate dibaryons only.

In a more realistic calculation, of course, one has to take into
account the interference of the dibaryon excitation mechanisms
with the background processes, mainly the $t$-channel
$\Delta\Delta$ and $NR$ excitation processes. However, one can see
already now that the dibaryon model, even in its simplest form
presented here, describes the data on $pp \to pp \pi^0 \pi^0$
reaction in a rather broad energy range not worse and even better
than the conventional model based on $t$-channel excitation of
hadronic resonances.

In the recent work of the WASA@COSY Collaboration~\cite{Adl14-pn},
the total cross section of a similar reaction $pn \to pn \pi^0
\pi^0$ in the GeV region has been measured. Experiment clearly
shows an enhancement due to the isoscalar resonance
$\mathcal{D}_{03}(2380)$ production at energies $T_p \simeq 1.1$
GeV. Therefore, it seems quite natural that the isovector
dibaryons $\mathcal{D}_{13}^-(2220)$ and $\mathcal{D}_{14}(2430)$
can be manifested in the reaction $pp \to pp \pi^0 \pi^0$ at
relevant energies (though the isovector resonance peaks will be
smeared in comparison to a more pronounced isoscalar peak due to
the larger widths of isovector resonances).

\begin{figure}[!ht]
\begin{center}
\resizebox{0.6\columnwidth}{!}{\includegraphics{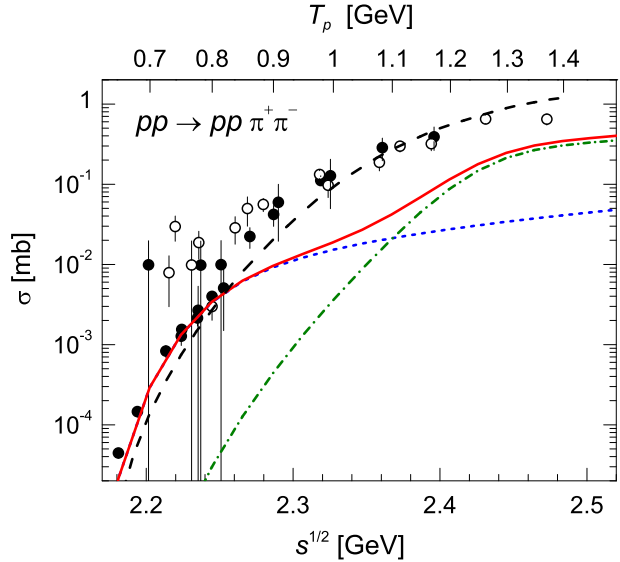}}
\end{center}
\caption{(Color online) Total cross section of the reaction $pp
\to pp \pi^+ \pi^-$. The meaning of theoretical curves is the same
as in Fig.~\ref{fig11a}. The dibaryon model calculation (solid
line) includes the isoscalar $\pi^+ \pi^-$ channel only. The
experimental data are shown by filled and open circles (see
references in~\cite{Cao10}).} \label{fig11b}
\end{figure}

We also calculated the contribution of the same dibaryon
resonances $\mathcal{D}_{13}^-(2220)$ and $\mathcal{D}_{14}(2430)$
(decaying with emission of an isoscalar $\pi\pi$ pair) to the
total cross section of the reaction $pp \to pp \pi^+\pi^-$. This
contribution, apart from isospin violation due to the mass
difference between the neutral and charged pions, would be just
twice the contribution to the reaction $pp \to pp \pi^0 \pi^0$.
The results of our calculation together with experimental data and
the Beijing model predictions are shown in Fig.~\ref{fig11b}.
Here, contrary to the reaction  $pp \to pp \pi^0 \pi^0$, the
dibaryon contributions are rather small everywhere except for the
near-threshold region. This result is not surprising given the
fact that the isovector $\pi\pi$ pairs should be produced with a
high probability in this reaction, however, their production is
suppressed near the threshold due to Bose symmetry (isovector
$\pi\pi$ pairs can be in odd partial waves only). The isovector
dipion production is described well by the conventional
meson-exchange models, as is also seen from Fig.~\ref{fig11b}.
Thus, one could expect manifestation of the intermediate dibaryons
in processes where the isoscalar pion-pair production dominates,
since the dibaryon in such cases can emit an intermediate
$\sigma$-meson which enhances the probability of such a resonance
mechanism. The signal of the near-threshold $\sigma$-meson
production can also be seen in the $\pi\pi$ invariant-mass
distribution for the reaction $pp \to pp \pi^0 \pi^0$ which is
concentrated near threshold and cannot be described by the
conventional meson-exchange models (see~\cite{Cao10}). This
feature can be regarded as some kind of the ABC effect in $pp$
collisions: though there is no significant enhancement over the
phase-space distribution (as in the $pn \to d (\pi\pi)_0$
reaction), there is a noticeable near-threshold enhancement over
the conventional model calculations (see also Sec.~\ref{pppn}). We
should also add here that the essential flatness of the angular
distributions in the reaction $pp \to pp \pi^0 \pi^0$ does not
contradict the interpretation of this reaction in terms of
intermediate dibaryons. Although the dibaryon resonances
considered here have high angular momenta, the integration over
the four-particle phase space and also interference with
background processes can flatten the observed angular
distributions. Indeed, as was shown in Ref.~\cite{Skorodko11}, the
angular distributions in the $pp \to pp \pi^0 \pi^0$ reaction in
case of the limited phase space (${}^2$He scenario) are much more
anisotropic than for the full four-particle phase space.

At the end of this section, it is interesting to note that some
attempts were made recently to modify the conventional Valencia
model~\cite{Alvarez98} in order to describe the numerous new
data~\cite{Skorodko11} on $2\pi$ production in $pp$ collisions. It
was found~\cite{Skorodko11} that for reproducing the basic
features of the total and differential cross sections in the
reaction $pp \to pp \pi^0\pi^0$, it is necessary to reduce the
$\rho$-meson exchange in the original model~\cite{Alvarez98} by an
order of magnitude (i.e., almost remove it). Furthermore, the
meson-baryon vertex form factors were dropped in this modified
version of the model. In fact, such a modification corresponds to
the account of the pion-exchange only, with the cut-off parameter
$\Lambda = \infty$. Although this model is hardly consistent with
the actual physical picture, it describes the data very
well~\cite{Skorodko11}. One could see here an interesting parallel
with the above-cited work~\cite{Ericson82}, where the deuteron
properties were accurately described in a model including only
pion exchange with $\Lambda = \infty$, as well as with
Ref.~\cite{Brack77}, where the total $\pi^+ d \to pp$ cross
section in a broad energy range was shown to be described
reasonably (though with incorrect normalization) within the same
model. All these observations might probably be related to the
general principle of continuity between hadron and quark d.o.f.

Thus, we have demonstrated that the one- and two-pion production
processes in $NN$ collisions can be consistently described in the
model involving excitation of intermediate dibaryon resonances
with realistic parameters. It has also been shown that the
dibaryon parameters used do not contradict the empirical data on
elastic $NN$ and $\pi d$ scattering.

To further clarify the relative role of the resonance and
background contributions to the one- and two-pion production
reactions in the GeV region, the more detailed knowledge of the
basic dibaryon parameters, along with independent confirmation of
the soft cut-off parameters in traditional meson-exchange
mechanisms is required. However, as will be shown below, even at
the present stage of our knowledge, an analysis of the inner
structure and possible decay modes of intermediate dibaryons allow
to give a qualitative explanation for some important experimental
observations which find no obvious explanation within the
conventional models.

\section{Two-pion production and dibaryon spectroscopy}
\label{dibspec}

In this section, we analyze the large differences between two-pion
production cross sections in $pn$ and $pp$ collisions in the
energy region $T_p \sim 1$ GeV in terms of intermediate dibaryon
resonances and their spectra. In fact, the total cross section for
production of the scalar-isoscalar pion pairs, i.e., $\pi^0\pi^0$
or $(\pi^+\pi^-)_0$, in $pn$ collisions at energies $T_p =
1.1$--$1.2$~GeV was found~\cite{Adl14-pn,Skorodko11} to be an
order of magnitude higher than that in $pp$ collisions. This
difference was interpreted~\cite{Adl14-pn} as a consequence of the
isoscalar dibaryon $\mathcal{D}_{03}(2380)$ excitation which
occurs in $pn$ collisions only. It is important to emphasize that
\emph{elastic cross sections} for $np$ and $pp$ scattering are, on
the contrary, very close to each other in the same energy region.
Furthermore, the $\pi\pi$ invariant mass distribution in the
reaction $pn \to d (\pi\pi)_0$ exhibits a pronounced
near-threshold enhancement (the ABC
effect)~\cite{Adl11,Adl13-iso}, whereas such an enhancement in the
reaction with similar kinematics $pp \to pp(^1S_0) \pi^0\pi^0 $
turns out to be very modest, if present at all~\cite{Skorodko11}.

In our previous work~\cite{PRC13}, an abnormally high yield of the
near-threshold scalar-isoscalar pion pairs observed in $pn$
collisions was quantitatively interpreted as a result of
constructive interference between two mechanisms of the
$\mathcal{D}_{03}(2380)$ resonance decay: a direct decay to the
final deuteron with emission of a light scalar $\sigma$ meson
(accompanied by a partial chiral symmetry restoration in a highly
excited dibaryon state~\cite{PRC13}) and two consecutive one-pion
decays via an intermediate isovector dibaryon
$\mathcal{D}_{12}(2150)$ production. While the latter mechanism of
sequential decay gives the rather uniform $M_{\pi\pi}$
distribution, the mechanism of the $\sigma$-meson emission, though
having a very small branching ratio, is highly concentrated near
the two-pion threshold.

If we now suppose (see the previous section) that two-pion
production in $pp$ collisions in the GeV region occurs also to a
large extent via intermediate (isovector) dibaryons formation,
then it would be interesting to investigate the reasons for
presence of the large ABC effect in $pn$ and its almost absence in
$pp$ collisions from this point of view. For this purpose, it is
important first to establish the relationship between isoscalar
and isovector dibaryons, as well as the impact of their possible
quark structure on the probability of two-pion production.

\subsection{Quark-cluster model for dibaryons}
\label{qcm}

The parameters (masses and total widths) for the main dibaryon
candidates determined from experiments and predicted
theoretically~\cite{Dyson64} on the basis of SU(6) symmetry are
summarized in Table II. Dibaryon candidates are presented in the
Table in order of decreasing experimental evidence. Thus, there is
solid experimental evidence for $\mathcal{D}_{01}(1.88)$ (the
deuteron), $\mathcal{D}_{10}(1.88)$ (the singlet deuteron) and
$\mathcal{D}_{03}(2.38)$, disputable evidence for
$\mathcal{D}_{12}(2.15)$, $\mathcal{D}_{13}^{-}(2.22)$ and
$\mathcal{D}_{12}^-(2.18)$\footnote{Note that this isovector
dibaryon has been observed just recently by the ANKE Collaboration
in the reaction $pp \to \{pp\}_s\pi^0$~\cite{Tsirkov15}.}, and
only experimental hints for $\mathcal{D}_{14}(2.43)$,
$\mathcal{D}_{15}^{-}(2.70)$ and $\mathcal{D}_{16}(2.90)$. In
quoting the parameters of isovector dibaryons in Table II we
follow mainly the summary paper of Yokosawa~\cite{Yokosawa90} and
also give references to some original papers concerning these
resonances. Two theoretically predicted dibaryons with high
isospins~\cite{Dyson64} which have not yet been seen
experimentally (see, however,~\cite{Clem14}) are also presented in
Table II.

%\begin{table*}[!ht]
%\flushleft {\caption{Average parameters of dibaryon resonances
%found from experiments (see~\cite{Yokosawa90} and references
%therein) in comparison with theoretical predictions~\cite{Dyson64}
%given in the last column.}}
%
%\begin{tabular*}{1.0\textwidth}{@{\extracolsep{\fill}}cccccc}\hline
%$\mathcal{D}$ & $I(J^P)$ & ${}^{2S+1}L_J (NN)$ & $M_D^{\rm exp}$
%[GeV] & $\Gamma_D$ [MeV] & $M_D^{\rm SU(6)}$ [GeV]
%\\ \hline $\mathcal{D}_{01}$ $(d)$ & $0(1^+)$ & ${}^3S_1$ & 1.88 &
%0 & 1.88 \\ $\mathcal{D}_{03}$ $(d^*)$ & $0(3^+)$ & ${}^3D_3$ &
%$\simeq 2.38$ &
%$\simeq 70$ & 2.35 \\ $\mathcal{D}_{10}$ & $1(0^+)$ & ${}^1S_0$ & 1.88 & $\simeq 0$ & 1.88 \\
%$\mathcal{D}_{12}$ & $1(2^+)$ & ${}^1D_2$ & $\simeq 2.15$ & $\simeq 120$ & 2.16 \\
%$\mathcal{D}_{13}^-$ & $1(3^-)$ & ${}^3F_3$ & $\simeq 2.22$ &
%$\simeq 150$ &
%--- \\ $\mathcal{D}_{14}$ & $1(4^+)$ & ${}^1G_4$ & $\simeq 2.43$ & $\simeq 150$ &
%--- \\ $\mathcal{D}_{15}^-$ & $1(5^-)$ & ${}^3H_5$ & $\simeq 2.70$ & $\simeq 150$ &
%--- \\ $\mathcal{D}_{16}$ & $1(6^+)$ & ${}^1I_6$ & $\simeq 2.90$ & $\simeq 150$ &
%--- \\ $\mathcal{D}_{21}$ & $2(1^+)$ & --- & ? & ? &
%2.16 \\ $\mathcal{D}_{30}$ & $3(0^+)$ & --- & ? & ? & 2.35
%\\ \hline
%\end{tabular*}
%\end{table*}

\begin{table*}[!ht]
\flushleft {\caption{Parameters of dibaryon resonances (in GeV)
found from experiments in comparison with theoretical
predictions~\cite{Dyson64} (the last column).}}

\begin{tabular*}{1.0\textwidth}{@{\extracolsep{\fill}}ccccccc}\hline
$\mathcal{D}_{IJ}$ & $I(J^P)$ & ${}^{2S+1}L_J^{(NN)}$ & $M_D^{\rm
exp}$ & $\Gamma_D^{\rm exp}$ & Refs. & $M_D^{\rm SU(6)}$
\\ \hline $\mathcal{D}_{01}(1.88)$ & $0(1^+)$ & ${}^3S_1$ & 1.88 &
0 & & 1.88 \\ $\mathcal{D}_{10}(1.88)$ & $1(0^+)$ & ${}^1S_0$ &
1.88 & $\simeq 0$ & & 1.88 \\ $\mathcal{D}_{03}(2.38)$ & $0(3^+)$
& ${}^3D_3$ & $2.38\!\pm\!0.01$ & $0.08\!\pm\!0.01$ &
\cite{Adl11,Adl14-el} & 2.35 \\
$\mathcal{D}_{12}(2.15)$ & $1(2^+)$ & ${}^1D_2$ & $2.14$--$2.17$ & $0.08$--$0.14$ & \cite{Auer78,Hoshizaki93,Bhandari81} & 2.16 \\
$\mathcal{D}_{13}^-(2.22)$ & $1(3^-)$ & ${}^3F_3$ & $2.20$--$2.25$
& $0.1$--$0.2$ & \cite{Auer82,Hoshizaki78,Bhandari81} &
--- \\
$\mathcal{D}_{12}^-(2.18)$ & $1(2^-)$ & ${}^3P_2$ & $2.17$--$2.20$
& $0.1$--$0.2$ & \cite{Arndt92,Yokosawa90} &
--- \\
$\mathcal{D}_{14}(2.43)$ & $1(4^+)$ & ${}^1G_4$ & $2.43$--$2.50$ &
$\simeq 0.15$ & \cite{Auer78,Auer82,Yokosawa90} &
--- \\ $\mathcal{D}_{15}^-(2.70)$ & $1(5^-)$ & ${}^3H_5$ & $2.70\!\pm\!0.1$ & $\simeq 0.15$
& \cite{Auer82,Auer89,Yokosawa90} &
--- \\ $\mathcal{D}_{16}(2.90)$ & $1(6^+)$ & ${}^1I_6$ & $2.90\!\pm\!0.1$ & $\simeq 0.15$
& \cite{Auer82,Auer89,Yokosawa90} &
--- \\ $\mathcal{D}_{21}$ & $2(1^+)$ & --- & ? & ? & &
2.16 \\ $\mathcal{D}_{30}$ & $3(0^+)$ & --- & ? & ? & & 2.35
\\ \hline
\end{tabular*}
\end{table*}

Although the isovector dibaryons have not reached the clear-cut
experimental evidence up to now, they were predicted by a number
of theoretical QCD-inspired models. Regarding the possible quark
structure of these dibaryons, one can follow theoretical arguments
and the respective models developed by the
Nijmegen~\cite{Mulders80} and ITEP~\cite{Kondr87} groups, based
partly on estimates for the masses of multiquark clusters obtained
in the MIT bag model. According to these models, the isovector
dibaryons with $J^P = 2^+, 3^-, 4^+, 5^-, \ldots$, observed in
$\vec{p} + \vec{p}$ scattering as very inelastic resonances, have
the two-cluster quark structure $[q^4-q^2]$, i.e., consist of a
tetraquark $q^4$ and a diquark $q^2$ connected by a colored QCD
string. The tetraquark with a mass $M(q^4) = 1.05$--$1.15$ GeV has
the quantum numbers\footnote {We use the letters $S$ and $T$ for
the spin and isospin of the multiquark clusters, in accordance
with notations adopted in~\cite{Kondr87}.} $(S = 1, T = 0)$, while
the diquark here is an axial one, i.e., with quantum numbers $(S =
T = 1)$ and a mass $M(q^2_A) = 450$--$550$~MeV. In general,
because the whole dibaryon states are colorless, while the quark
clusters $q^4$ and $q^2$ as well as the string between them are
colored objects, one is dealing in this case with a ``hidden
color'' (first predicted by Brodsky et
al.~\cite{Brodsky83,Brodsky86}; see also the recent
paper~\cite{Bash13}). So, the isovector and isoscalar dibaryons
considered here can be classified as the typical hidden-color
objects.

Next, according to~\cite{Kondr87}, the observed series of
isovector dibaryons lies on a relativistic Regge trajectory which
describes rotational excitations of a relativistic string
connecting two multiquark clusters. An important contribution to
dibaryon masses is also given by the spin-orbit interaction
between the quark clusters and the rotating string~\cite{Kondr87}.
In this case, the trajectory of isovector dibaryon states on the
graph $[J, M_D^2]$ (where $J$ is a total dibaryon angular
momentum) does not necessarily correspond to a straight line.

On the other hand, for relatively low energies of the rotational
excitation $E^* = \Delta M \ll M_0$, where $M_0$ is a mass of the
lowest rotational state, one can use the non-relativistic
description of the rotational excitations in a clustered system
$[q^4 - q^2]$ with an orbital angular momentum $L$ between the
quark clusters. For successively increasing values of $L =
0,1,2,3,\ldots$, one obtains the respective isovector dibaryons
with alternating parities:\footnote{We omit the ``+'' superscript
for the positive-parity states.} $\mathcal{D}_{12}(2.15)$,
$\mathcal{D}_{13}^{-}(2.22)$, $\mathcal{D}_{14}(2.43)$,
$\mathcal{D}_{15}^{-}(2.7)$, $\ldots$ In this case, the rotational
band of isovector dibaryons can be described by a simple
non-relativistic formula (corresponding to the model of a rigid
rotor) for the rotational states in the $[q^4 - q^2]$ system, with
an additional term $M_{LS}$ due to the spin-orbit interaction:
\begin{equation}
M_D(L) = M_0 + \beta \frac{\hbar^2}{2\mathcal{I}}L(L+1) + M_{LS},
\end{equation}
where $\mathcal{I}$ is a moment of inertia for the rotating
quark-cluster system and the constant $\beta$ takes into account
the kinetic energy of the rotating string itself.

\begin{figure}[!ht]
\begin{center}
\resizebox{0.65\columnwidth}{!}{\includegraphics{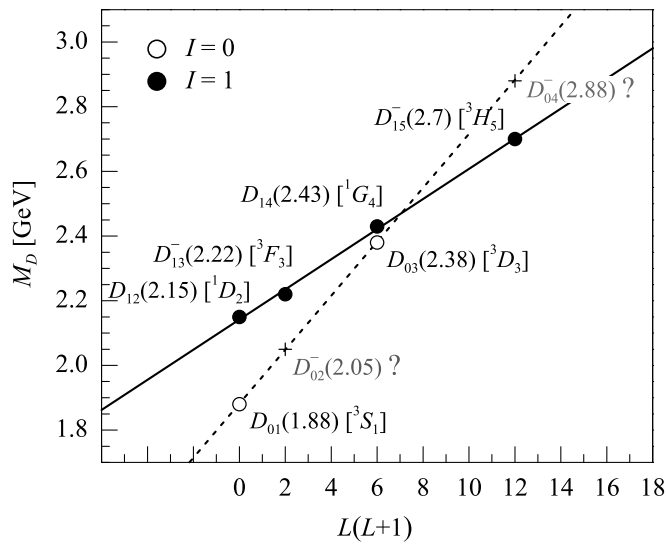}}
\end{center}
\caption{Dependence of the masses of known isovector and isoscalar
dibaryon (six-quark) states on the square of the orbital angular
momentum $L(L + 1)$ between the tetraquark ($q^4$) and diquark
($q^2$) clusters. The dibaryon masses are given in parentheses,
and the dominant partial waves of the $NN$ system, in which the
respective dibaryons can be excited, are shown in square brackets.
The positions of hypothetical negative-parity isoscalar dibaryons
uncoupled from the $NN$ channel are marked by crosses.}
\label{fig12}
\end{figure}

Dependence of the isovector dibaryon masses $M_D$ on the quantity
$L(L + 1)$ is shown in Fig.~\ref{fig12}. It is clearly seen that
the masses of the known isovector dibaryons are well fitted into a
straight line. This gives a strong argument in favor of the above
quark-cluster structure of dibaryons, with multiquark clusters at
the ends of the rotating colored string, which is well described,
at least for a few lowest states, by the non-relativistic rigid
rotor model. In this case, the correction due to the spin-orbit
interaction apparently does not lead to any significant deviation
from a straight line on a graph $[L(L+1), M_D]$.

It is interesting to note that in some previous works (see,
e.g.,~\cite{MacGregor79}) the same isovector dibaryons were
considered as lying on a rotational band in the $NN$ system, in
the spirit of rotational bands in nuclear physics. It was
found~\cite{MacGregor79} that the trajectory of isovector
dibaryons on the graph $[L_{NN}(L_{NN} +1), M_D]$, where $L_{NN} =
L+2 = J$, is also rather close to a straight line. However, since
the dibaryons are known to be highly inelastic in the $NN$
channel, the description in terms of quark clusters $[q^4 - q^2]$
rather than $[N - N]$ looks to be more appropriate.

In Fig.~\ref{fig12}, the two known isoscalar dibaryons, i.e., the
deuteron dibaryon $\mathcal{D}_{01}(1.88)$ and the resonance
$\mathcal{D}_{03}(2.38)$, are also shown. These dibaryons are
actually different from their isovector analogues with the same
orbital angular momentum $L$, i.e., $\mathcal{D}_{12}(2.15)$ and
$\mathcal{D}_{14}(2.43)$, respectively, by replacing the axial
diquark by a scalar one, having quantum numbers $(S=T=0)$.
Accordingly, the spin-orbit interaction between the diquark and
the colored string which shifts down the isovector dibaryon masses
with $L > 0$~\cite{Kondr87}, should be turned off for their
isoscalar partners. Then, assuming the isoscalar dibaryon series
to be described by the above rigid rotor model, and drawing a
straight line connecting the deuteron with the $\mathcal{D}_{03}$
resonance, one can predict the existence of two isoscalar
dibaryons of negative parity with masses lower than 3 GeV, namely,
the $\mathcal{D}_{02}^-$(2.05) and $\mathcal{D}_{04}^-$(2.88).
Since these dibaryons correspond to odd values of $L$ and thus $L
+ S + I = L + 1$ is an even number, they should be uncoupled from
the $NN$ channel.

Actually, the dibaryon $\mathcal{D}_{02}^-$ (or $d'$)
corresponding to $L = 1$ was predicted previously by both the
Nijmegen~\cite{Mulders80} and ITEP~\cite{Kondr87} groups.
Moreover, in the ITEP model~\cite{Kondr87}, the $d'$ dibaryon was
predicted to have the same mass $M_{d'} \simeq 2.05$ GeV as we got
here. Unfortunately, experimental searches for this dibaryon in
previous years have not led to any unambiguous and convincing
conclusions about its existence~\cite{Draeger00,Brodowski02}. The
second negative-parity isoscalar dibaryon $\mathcal{D}_{04}^-$
(which one might call $d'^*$) with a mass $M_{d'^*} \simeq 2.88$
GeV, as far as we know, is predicted \emph{for the first time} in
the present paper. So, the renewal of experimental search for
these isoscalar dibaryons, as well as finding new confirmations
and refining the parameters of the currently known isovector
dibaryons seem to be the important further steps towards
determining the true structure of the resonance $6q$ states.

According to the ITEP model~\cite{Kondr87}, just the two-cluster
structure with sufficiently separated multiquark clusters allows
existence of the relatively long-lived $6q$ configurations with
total widths $\Gamma_{6q} \simeq 100$--$150$~MeV, that is, of the
same order as the $\Delta$-isobar width. It is important to add
here that the same effect of a $6q$-system clustering into a
tetraquark and a diquark was found also in $S$-wave $NN$
interaction within the dibaryon model for nuclear
force~\cite{JPG01K,IJMP02K}. Thus, clustering the dibaryons in the
deuteron or singlet deuteron into a tetraquark $q^4 (S = 1, T =
0)$ or $q^4 (S = 0, T = 1)$, respectively, and a scalar diquark
$q^2 (S' = T' = 0)$ is achieved by a two-quantum ($2\hbar\omega$)
excitation of a colored string with an orbital angular momentum $L
= 0$. This $2 \hbar \omega$ excitation (being a simple consequence
of the dominating $s^4p^2[42]$ symmetry in the $6q$
system~\cite{JPG01K}) gives rise to a node in the radial wave
function of the multiquark system, which corresponds exactly to
the two-cluster structure $[q^4 - q^2]$, although in this case the
quark clusters are in a relative $S$ wave. It is important to note
that the same picture for $S$-wave $NN$ interaction was found also
in a fully microscopic calculation of the $6q$ system in the
resonating group method~\cite{Stancu97,Bartz01}. The
authors~\cite{Stancu97,Bartz01} (see also~\cite{Faessler83}) found
that the two-cluster configuration $[q^4 - q^2]$ with a radial
node in its relative-motion wave function dominates the six-quark
wave function of the $NN$ system in $^3S_1$ and $^1S_0$ channels.
Thus, the clusterization of six-quark states seems to be a general
phenomenon providing the relatively long-lived intermediate
resonances in the $NN$ interaction. So, it might play an essential
role in the short-range nuclear force.

Given the above isovector and isoscalar rotational bands of
dibaryons, one can consider the transitions between different
dibaryon states via the meson emission. Thus, transitions between
the eigenstates of two different bands can occur naturally via a
pion emission changing a scalar diquark to an axial one and vice
versa. Transitions within the same band can occur most likely via
a string deexcitation through a light scalar ($\sigma$) meson
emission. As was shown above, such transitions can generally be
observed in one- or two-pion production processes in $NN$
collisions however interfering with conventional processes
involving intermediate baryonic resonances. The most ``clear''
case in this respect seems the pure isoscalar reaction $pn \to d
(\pi\pi)_0$.

Now, using the above model for dibaryon resonances and also the
results of Ref.~\cite{PRC13}, we consider the differences in the
cross sections for two-pion production in $pn$ and $pp$ collisions
at energies $T_p \sim 1$ GeV.

\subsection{Comparison of $2\pi$-production cross sections in $pn$ and $pp$ collisions}
\label{pppn}

Here, we focus mainly on presence of the pronounced near-threshold
enhancement (ABC effect) in the $M_{\pi\pi}$ spectrum in the
isoscalar reaction $pn \to d(\pi\pi)_0$ and the near absence of it
in the similar isovector reaction $pp \to pp({}^1S_0) (\pi\pi)_0$.
First of all, it is important to emphasize that, even in the
isoscalar $NN$ channel, a significant ABC effect is observed only
in case of the \emph{bound state} (the deuteron) formation in the
final $pn$ system. This fact was confirmed by the latest
measurements~\cite{Adl14-pn} for the reaction $pn \to pn \pi^0
\pi^0$, which revealed a strong $\mathcal{D}_{03}$ resonance
signal in the total cross section, but the very modest ABC
enhancement in the $M_{\pi\pi}$ spectrum. From the first sight, as
claimed in~\cite{Adl14-pn}, it may pose a problem for the
interpretation of the ABC effect~\cite{PRC13} as a consequence of
a light scalar $\sigma$ meson production within the process $pn
\to \mathcal{D}_{03} \to d + \sigma \to d + (\pi\pi)_0$. However,
one should bear in mind that the $M_{\pi\pi}$ distribution in
reaction $pn \to pn \pi^0 \pi^0$ is obtained by an integration
over the available invariant masses $M'_{pn}$ of the final $pn$
pair. Since the $\sigma$-meson is emitted in the $D$ wave from the
$\mathcal{D}_{03}$ decay, and this process is concentrated near
the $M_{\pi\pi}$ threshold even in case of the final
deuteron~\cite{PRC13}, the large centrifugal barrier will strongly
suppress the $\sigma$-meson emission at higher $M'_{pn}$.
Therefore, in the integrated $M_{\pi\pi}$ distribution, the
$\mathcal{D}_{03} \to pn + \sigma$ branch and thus the ABC
enhancement should be hardly visible, in accordance with
experimental data~\cite{Adl14-pn}. Besides that, the process
$\mathcal{D}_{03} \to d + \sigma$ is likely to be dynamically
selected from other final $pn$ configurations, since it represents
a direct transition between two discrete eigenstates of the $6q$
system. On the contrary, the process $\mathcal{D}_{03} \to
\mathcal{D}_{12} + \pi^0 \to pn \pi^0\pi^0$ will ``survive'' the
$M'_{pn}$ integration, since the decay of the resonance
$\mathcal{D}_{12}$ into $NN \pi$ channel is known to have a larger
probability than that into $d \pi$ channel. This may reflect the
fact that the isovector resonance $\mathcal{D}_{12}$ have a large
$N + \Delta$ component, with the rather weakly bound $N + \Delta$
system, unlike the deeply bound $\Delta + \Delta$ system in the
isoscalar $\mathcal{D}_{03}$ state. So, an intermediate $N +
\Delta$ state can give a large contribution to the
$\mathcal{D}_{12}$ decay into $NN \pi$ channel. A detailed
calculation is in order to test these qualitative considerations,
however, our purpose in this section is to give just a qualitative
insight into the differences between the observed cross sections
for $2 \pi$ production in different $NN$ channels.

Let's now turn to the $2 \pi$ production reactions in the
isovector $NN$ channel. If one interprets the isoscalar dipion
production in terms of intermediate dibaryons, then in case of
$pp$ collisions at energies $T_p \sim 1$ GeV, the two-pion
emission should proceed most likely through the decay of
intermediate isovector dibaryons\footnote{For the reader's
convenience, we denote here the isovector dibaryons by the
respective quantum numbers of the $NN$ channel.} $^3F_3(2220)$ and
$^1G_4(2430)$ (see Sec.~\ref{twopi}). It is easy to see however
that the direct transitions $\mathcal{D} \to pp({}^1S_0) +
\sigma$, where $\mathcal{D}$ is one of the above two dibaryons,
with emission of a scalar $\sigma$ meson, although possible in
case of a low $\sigma$ mass $m_{\sigma} \simeq 300$~MeV (which may
follow from chiral symmetry restoration in an excited
dibaryon~\cite{PRC13}), will be suppressed by the centrifugal
barrier, since the $\sigma$ meson has to be emitted from the
dibaryon $^3F_3$ or $^1G_4$ decay in $F$ or $G$ waves,
respectively. For comparison, in case of the isoscalar dibaryon
$\mathcal{D}_{03}(2380)$ decay into $d + \sigma$ channel, one has
approximately the same available phase space as for the decay
$^1G_4(2430)\to pp ({}^1S_0) + \sigma$ (with the same mass of the
$\sigma$ meson), but the $\sigma$ meson in first case is emitted
in $D$ wave, thus leading to a pronounced ABC enhancement in the
$2\pi$ invariant-mass spectrum in the $pn \to d(\pi\pi)_0$
reaction~\cite{PRC13}. If we take into account the large width of
the $^1G_4(2430)$ dibaryon, then it is also possible to consider
the decay $^1G_4(2430) \to {}^1D_2(2150) + \sigma$ with the
$D$-wave $\sigma$-meson emission, but this decay route will be
suppressed due to the small available phase space.

Furthermore, the probability for the above isovector dibaryons to
decay into the singlet deuteron ${}^1S_0(1880)$ with emission of a
scalar $\sigma$ meson should be additionally reduced because of
the quark structure of these dibaryon states. In the quark-cluster
model for dibaryons described above, the dibaryon component of the
singlet deuteron has the structure $[q^4 (S = 0, T = 1) - q^2 (S'
= T' = 0)]$, whereas the structure of dibaryons $^3F_3$ and
$^1G_4$ is $[q^4 (S = 1, T = 0) - q^2 (S' = T' = 1)]$. Therefore,
in two-pion emission from such isovector dibaryons, the one-pion
transition of an axial diquark into a scalar one, i.e., $q^2 (S' =
T' = 1) \to q^2 (S' = T' = 0)$, must be accompanied by a
simultaneous one-pion Gamow--Teller transition in the tetraquark,
i.e., $q^4 (S = 1, T = 0) \to q^4 (S = 0, T = 1)$. Hence the
generation of a tightly correlated scalar-isoscalar pion pair
under such conditions should be very unlikely. On the contrary,
the isoscalar dibaryon $\mathcal{D}_{03}(2380)$ and the dibaryon
component of the deuteron have the same quark-cluster structure
$[q^4 (S = 1, T = 0) - q^2 (S' = T' = 0)]$, thus the $\sigma$
meson is emitted here via a direct deexcitation of the colored
string from a rotational level $L = 2$ to the ground level $L =
0$, i.e., without rearrangement of the quark clusters themselves.

Thus, in general, emission of a light scalar $\sigma$ meson near
the $2\pi$ threshold which can explain the significant ABC
enhancement in the isoscalar $NN$ channel, appears to be
relatively suppressed in the isovector $NN$ channel. As a result,
one comes to a conclusion well confirmed by
experiments~\cite{Skorodko11}, that the ABC effect in the reaction
$pp \to pp \, (\pi\pi)_0$, including the limiting case $pp \to
pp({}^1S_0) \, (\pi\pi)_0$, is very small, if visible at all
(although it can still be manifested under certain kinematic
conditions~\cite{Dymov09}).

Given the above arguments, it is possible to understand
qualitatively the observed differences between two-pion production
cross sections in $pn$ and $pp$ collisions. In this respect, the
interpretation of $2\pi$ production in $NN$ collisions in terms of
generation of intermediate dibaryon resonances and their possible
decay modes with two-pion emission seems to be rather appropriate
and natural.

\section{Conclusions}
\label{concl}
In this work, we analyzed the contributions of
intermediate dibaryon resonances to the one- and two-pion
production processes in $NN$ collisions. Since these processes, in
contrast to elastic scattering, are always accompanied by a large
momentum transfer, i.e., involve the region of small inter-nucleon
distances, a very important role in description of such processes
is played by the short-range mechanisms of $NN$ interaction, based
on quark structure of interacting nucleons. In particular, in the
overlap region of two nucleons, the probability of generation of
the compact six-quark objects, i.e., dibaryon resonances, can
increase considerably. This conclusion is confirmed in the present
study by a comparative analysis of contributions of $s$-channel
dibaryon-formation and $t$-channel meson-exchange mechanisms to
the elastic $pp$ and $\pi^+ d$ scattering and the one-pion
production reaction $pp \to d\pi^+$. An even more pronounced
manifestation of intermediate dibaryon resonances is expected in
two-pion production reactions. In addition to the continuing study
of the isoscalar $0(3^+)$ resonance observed recently in two-pion
production in $pn$ collisions~\cite{Adl11}, we proposed searching
the signals of isovector dibaryons in $pp$ collisions.

However, as shown by numerous studies including the present work,
the contributions of short-range QCD mechanisms can often be
simulated rather accurately by the conventional meson-exchange
mechanisms (with appropriate parameter fitting). This fact can
explain the success of meson-exchange models in the description of
many hadronic and electromagnetic processes including those with
high momentum transfers. However, while the long-range part of
$NN$ interaction is described universally by $t$-channel
meson-exchange mechanisms (mainly by one- and two-pion exchange)
and poses no doubts, an effective description of its short-range
part requires introducing the specific mechanisms and careful
adjusting their parameters (mainly the vertex cut-offs) \emph{ad
hoc} to a particular process. These parameters entering the same
mechanisms should be changed to describe the different processes
and are often not consistent with microscopic predictions. As an
example, one can consider the large differences between the
short-range cut-off parameter in the $\pi N \Delta$ vertex, needed
to describe elastic $\pi N$ scattering, from parameters in the
same vertex used in realistic potential models for $NN$
interaction, in conventional models for one-pion production and
others. In other words, the description of short-range processes
in traditional meson-exchange models is not entirely consistent
and contains a number of inner contradictions (see discussion on
this issue in~\cite{Holinde92}).

%It is worth noting that one faces a some similar situation in the
%chiral perturbation theory ($\chi PT$)~\cite{EFT}, where the
%description (i.e., the parameter values) of the short-range $NN$
%interaction must be modified when one goes to next orders of the
%chiral expansion.

On the other hand, in effective description of the short-range QCD
mechanisms of $NN$ interaction by using dibaryon degrees of
freedom, the dibaryon parameters used in calculations of various
hadronic and nuclear processes are in a good agreement with each
other. It is important to realize that QCD-motivated dibaryon
mechanisms at small inter-nucleon distances do not contradict the
traditional meson-exchange picture at the large and intermediate
distances, but rather complement it. The dibaryon generation does
not actually contradict also the heavy-meson exchange, provided
the realistic (soft) cut-off parameters in the respective vertices
are used. However, concerning generation of the heavy vector
mesons $\rho$ and $\omega$ at short $NN$ distances, it seems more
natural to assume these mesons emerging from a unified meson cloud
of a six-quark object (dibaryon) than in $t$-channel exchange
between two isolated nucleons at distances $r_{NN} \sim 0.2$ fm,
where the quark cores of two nucleons are strongly overlapped (see
the detailed discussion in~\cite{PAN13}).

Thus, we believe that dibaryons are much more intriguing objects
than just multiquark exotics, which might be manifested under
specific experimental conditions. They seem to be a manifestation
of the fundamental properties of nonperturbative QCD, which drive
the $NN$ interaction at short distances and, in general, the
short-range correlations in nuclei. The quantitative verification
of this hypothesis requires further theoretical and experimental
research.

\section*{Acknowledgements} The work was done under partial financial support
from RFBR grants Nos.~12-02-00908 and 13-02-00399. M.N.P. also
appreciates support from Dynasty Foundation.

\section*{References}

\bibliography{mybibfile}

\end{document}